\documentclass[fleqn,usenatbib]{mnras}

\usepackage{newtxtext,newtxmath}

\usepackage[T1]{fontenc}
\usepackage{ae,aecompl}

\usepackage{graphicx}	
\usepackage{amsmath}	
\usepackage{amssymb}	
\usepackage{CJKutf8}

\newcommand{\nustar}{\textit{NuSTAR }}
\newcommand{\suzaku}{{\it Suzaku }}
\newcommand{\swift}{{\it Swift }}
\newcommand{\xmm}{{\it XMM-Newton }}

\newcommand{\green}{\textcolor{red}}

\title[Spectral Fitting of 1H0419$-$577]{A Relativistic Disc Reflection Model for 1H0419$-$577: Multi-Epoch Spectral Analysis with \xmm and \nustar}

\author[J. Jiang]{Jiachen Jiang\begin{CJK*}{UTF8}{gbsn}
(姜嘉陈)
\end{CJK*}$^{1}$\thanks{E-mail: jj447@cam.ac.uk}, Dominic J. Walton$^{1}$, Andrew C. Fabian$^{1}$ and\newauthor Michael L. Parker$^{2}$
\\
$^{1}$Institute of Astronomy, University of Cambridge, Madingley Road, CB3 0HA Cambridge, UK\\
$^{2}$European Space Agency (ESA), European Space Astronomy Centre (ESAC), E-28691 Villanueva de la Ca\~nada, Spain\\
}
\date{Accepted XXX. Received YYY; in original form ZZZ}

\pubyear{2018}

\begin{document}
\label{firstpage}
\pagerange{\pageref{firstpage}--\pageref{lastpage}}
\maketitle

\begin{abstract}
We present a detailed analysis of the spectral properties of the Seyfert 1 galaxy 1H0419$-$577, based on the archival \textit{XMM-Newton}, \nustar and simultaneous \swift observations taken between 2002-2015. All the observations show a broad emission line feature at the iron band. We demonstrate that the broad band spectral variability at different levels can be explained by the combination of light-bending effects in the vicinity of the central black hole plus a thin warm absorber. We obtain a black hole spin of $a>0.98$ by fitting the multi-epoch spectra with the relativistic disc reflection model. 1H0419$-$577 is accreting at $40\%$ of its Eddington limit and its X-ray band shows the hardest powerlaw continuum in the highest flux state, which was previously more commonly seen in AGNs with a low accretion rate (e.g. $L_{\rm x}/L_{\rm Edd}<10^{-2}$). The \nustar observation shows a cool coronal temperature of $kT=30^{+22}_{-7}$\,keV in the high flux state. 
\end{abstract}





\section{Introduction}

In General Relativity, the spacetime geometry around a black hole (BH) is described by the Kerr solution \citep{kerr96} assuming charge neutrality, where the black hole is characterized by its mass $M_{\rm BH}$ and its dimensionless spin parameter $a_*=a/M_{\rm BH}=Jc/GM_{\rm BH}^2$ (where $J$ is the angular momentum). The dimensionless spin parameter affects the spacetime around BHs in either X-ray Binaries ($M_{\rm BH}\approx5-20M_{\odot}$), or in Active Galactic Nuclei (AGNs) ($M_{\rm BH}\approx10^{6-10}M_{\odot}$), in a similar behavior once the distance, timescale and luminosity scaled up by the corresponding BH mass \citep{mchardy06,walton12b}. BH spin is important for a variety of areas of BH research. For example, it has been suggested that the BH spin may be involved in powering jets \citep{blandford77}. From the fundamental physics point of view, possible generic deviations to the Kerr metric tend to have similar effects on the electromagnetic spectrum as the spin parameter \citep[e.g.][]{johannsen10,jiang15}, although no clear evidence of large deviations has been seen so far \citep[e.g.][]{bambi18}. 

One method of measuring black hole spins is using relativistic reflection spectroscopy. This approach has been applied to both the stellar-mass BHs in X-ray binaries and the supermassive black holes (SMBH) in AGNs. The central assumption of this method is that the inner edge of the accretion disc around the central BH is located at the inner-most stable circular orbit (ISCO). The measurement of the spin is based on the simple positive correlation between the spin and ISCO: $R_{\rm ISCO}=6R_{g}$ for Schwarzschild black hole ($a_*=0$) and decreases to $R_{\rm ISCO}=1R_{g}$ for a maximumly spinning Kerr black hole \citep[$a_*\approx1$;][]{bardeen72}. The accretion disc is irradiated by a high temperature compact structure external to the disc, producing a reflected component called the disc reflection spectrum. This high temperature structure is called the corona. The disc reflection spectrum consists of broad emission lines and a Compton back-scattered continuum. The emission line features are broadened by strong Doppler effects and gravitational redshifts in the vicinity of BHs. The most prominent broad emission line feature is the broad iron K$\alpha$ emission line, which has now been seen in various AGNs, such as MCG-6-30-15 \citep[e.g.][]{tanaka95,wilms01,reflection3,marinucci}, 1H0707-495 \citep[e.g.][]{fabian3,1h0707}, Mrk 335 \citep[e.g.][]{larsson08,gallo13,walton13,parker14}, NGC 3783 \citep[e.g.][]{brenneman11,reis12}, NGC 1365 \citep[e.g.][]{risaliti13,walton14}, IRAS 00521-7054 \citep[e.g.][Walton et al. in prep]{tan12,ricci14}, and Swift J2127.4+5654 \citep[e.g.][]{miniutti09,marinucci14}.

1H0419$-$577 \cite[z=0.104,][]{thomas98} is confirmed to be a Seyfert 1 galaxy \citep{guainazzi98}. The centre of 1H0419$-$577 hosts a supermassive black hole with $M_{\rm BH}=1.3\times10^{8}M_{\odot}$ by measuring its H$\beta$ line width \citep[$FWHM=2580\pm200$\,km\,s$^{-1}$][]{grupe10}. 

A highly variable soft band (<2~keV) has been found in 1H0419$-$577 with \textit{ROSAT} \citep[e.g.][]{guainazzi98}. Also a variable power-law continuum emission in the high energy band was noticed in later \xmm observations \citep{pounds04a,pounds04b,pounds05}. A long \xmm observation in 2010 shows a thin, lowly ionized warm absorber in its high resolution grating spectrum \citep{gesu13}. By fitting the \xmm spectrum at an extreme low flux state with the relativistic reflection model, \citet{fabian05} obtained a disc inner radius of $r_{\rm in}<2 r_{\rm g}$, indicating a black hole spin of $a_{*}>0.95$. Similarly, \citet{walton13} obtained a black hole spin measurement of $a_*>0.88$ by analyzing its \suzaku observation which also shows a broad iron K$\alpha$ emission line at a high X-ray flux state. No significant evidence of a fast outflow has been found in 1H0419$-$577 \citep{tombesi1}. A possible low coronal temperature has been reported by fitting the hard X-ray \nustar spectra with an absorption model \citep{turner18}. However \citet{parker15} studied the principal components in the X-ray variability of 1H0419$-$577, finding that the suppression of the primary component at the iron band and low energies cannot be explained by variable absorption models. 

In this work, we study the inner BH accretion disc in 1H0419$-$577 by analyzing its different X-ray flux states captured by all the archival \xmm and \nustar observations and try to explain the spectral variability with light-bending effects. A robust measurement of the relativistic parameters, including the black hole spin $a_*$ and the disc viewing angle $i$, is obtained by conducting a multi-epoch spectral analysis and running Markov chain Monte Carlo (MCMC) chains. 


\section{Data Reduction}

1H0419$-$577 was observed by a series of 12-18\,ks \xmm short looks in 2002 and 2003. Two longer \xmm observations were obtained in 2010 with a total exposure of $\approx160$\,ks. Note that during one of the \xmm observations (Obs ID 0148000701), the EPIC-pn exposure is dominated by a high flaring particle background, and thus ignored for this work. The hard X-ray satellite \nustar observed 1H0419$-$577 for a net exposure of 170\,ks in 2015 with a simultaneous 2\,ks \swift short look. A list of all the observations considered in this work is shown in Table \ref{tab_obs}. In this section, we introduce our data reduction process for all these observations.

\begin{table}
\caption{The list of the observations analyzed in this work. The exposure time for the \xmm observations is clean of the time intervals of high flaring particle background. LW: large window mode; SW: small window mode; PC: photon counting mode.}
\label{tab_obs}
\begin{tabular}{lllll}
\hline\hline
Satellite & Obs ID       & Start Date & Exp(ks) & Mode \\
\hline
XMM (pn)       & 0148000201  & 2002-09-25 & 11.5    & LW   \\
          & 0148000301  & 2002-12-27 & 0.3     & LW   \\
          & 0148000401  & 2003-03-30 & 11.0    & LW   \\
          & 0148000501  & 2003-06-25 & 5.8     & LW   \\
          & 0148000601  & 2003-09-16 & 11.3    & LW   \\
          & 0604720301  & 2010-05-30 & 71.0    & SW   \\
          & 0604720401  & 2010-05-28 & 42.3    & SW   \\
\hline
NuSTAR        & 60101039002 & 2015-06-03 & 170     & -    \\
\hline
Swift (XRT)    & 00081695001 & 2015-06-03 & 2.2     & PC     \\
\hline
\end{tabular}
\end{table}

\subsection{\xmm data reduction}

The \xmm data are reduced using the \xmm Science Analysis System (SAS) V15.0.0 and calibration files (ccf) v.20160201. We only consider EPIC-pn data in this work. The tool EPPROC is used to create clean calibrated event lists. We filter the data for flaring particle background and the background-dominated intervals are defined as the intervals where the single event count rate in the 10--12~keV band larger than 0.4 counts~s$^{-1}$. The spectra are extracted using the tool EVSELECT, selecting both the single and double events in a circular region with a radius of 35\,arcsec. Background regions are chosen on the same chip in the region to avoid any issues due to background  Cu~K emission lines from the electronic circuits on the back side of the detector\footnote{See following link for more details. https://xmm-tools.cosmos.esa.int/external/xmm\_user\_support/documentation/uhb /epicintbkgd.html}. A circular background region with a radius of 50\,arcsec in a source-free region near the source is used for observations in the small window (SW) mode (obsID 0604720301-0604720401). None of \xmm observations suffer from obvious pile-up effects. The ARFGEN and RMFGEN tasks are used to generate redistribution matrix files and auxiliary response files. We concentrate on the EPIC-pn spectra in the energy range of 0.5--10~keV for multi-epoch spectral analysis, due to its high effective photon collecting area. The \xmm spectra are grouped to have a minimum number of 50 counts per bin. 

\subsection{\nustar data reduction}

1H0419$-$577 was observed by the \nustar satellite in 2015 for $\approx170$\,ks. The \nustar data are reduced using the standard pipeline NUPIPELINE V0.4.6 and instrumental responses from \nustar caldb V20171002. We extract the source spectra from circular regions with radii of 100\,arcsec, and the background spectra from nearby circular regions on the same chip. The tool NUPRODUCTS is used for this purpose. The 3-78\,keV band is considered for both FPMA and FPMB spectra. The FPM spectra are grouped to have a minimum number of 50 counts per bin.

\subsection{\swift data reduction}

One short \swift observation was taken during the \nustar observation in 2015. The XRT was operated in the photon counting (PC) mode for $\approx 2$\,ks. The calibration file version used is 20160609. The source spectrum is extracted from a circular region with a radius of 50 arcsec and the background spectrum is extracted from a circular region with a radius of 200 arcsec nearby. The spectrum is binned to have a minimum count of 50 per bin. The averaged count rate is only 0.36 counts\,s$^{-1}$ in the 0.5--6\,keV band, which is lower than the pile-up threshold.  

\begin{figure}
\centering
\includegraphics[width=\hsize]{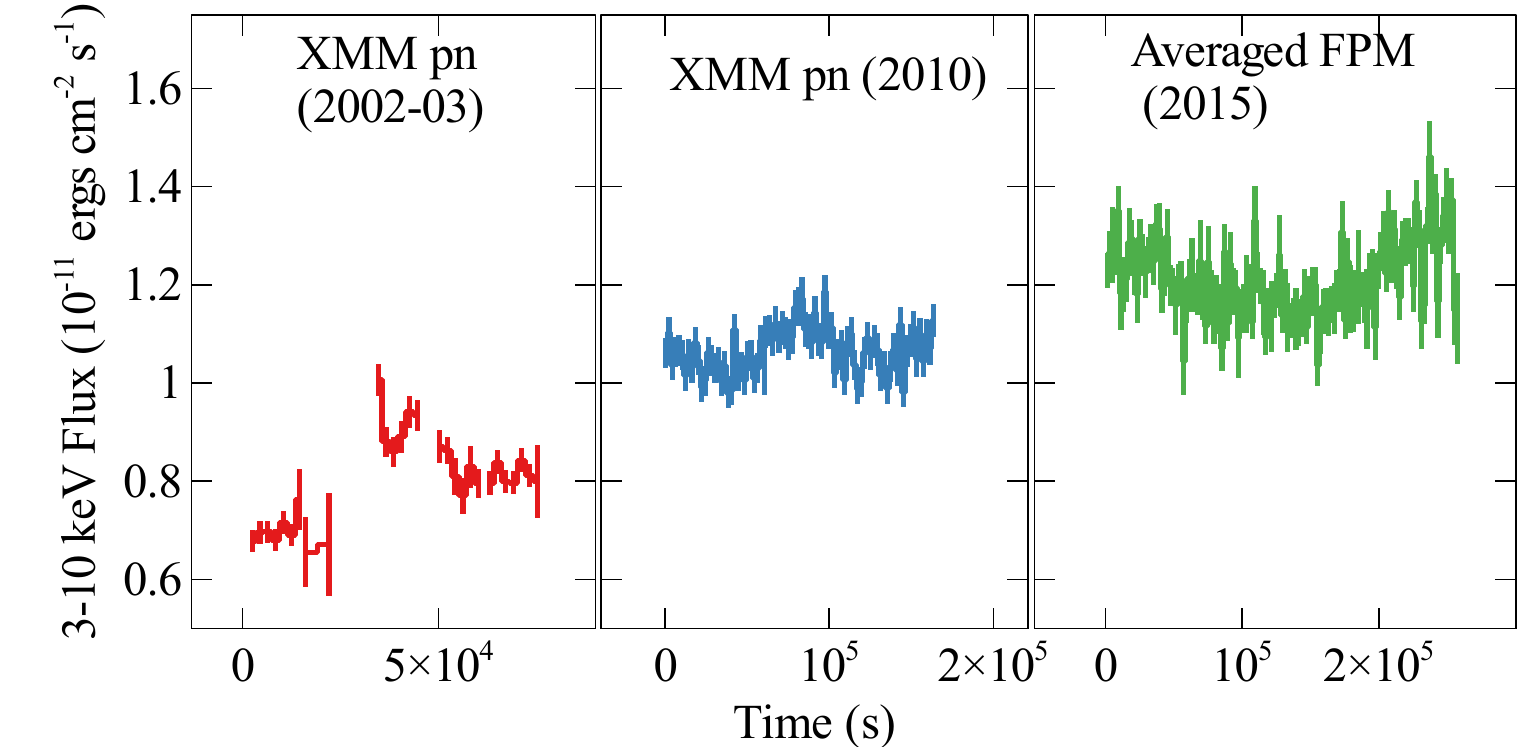}
\caption{The 3-10\,keV light curves of the archival \xmm (red and blue) and \nustar (green) 1H0419$-$577 observations in 2~ks bin. The first series of \xmm observations in 2002-2003 (red) show that the source is at the lowest flux state. The second 160~ks \xmm observation in 2010 (blue) shows an intermediate flux state. The \nustar (green) observations show a higher flux state.}
\label{pic_lc}
\end{figure}

\section{Broad Band Spectral Analysis} 

The 3--10\,keV band lightcurves of all the observations in 2\,ks bin are shown in Fig. \ref{pic_lc}. The first series of \xmm observations in 2002 and 2003 was taken when the source is at the lowest flux (LF, red in figures) level among the observations analyzed in this work; a longer \xmm look of the source in 2010 show a middle flux (MF, blue in figures) state; the \nustar observation shows a high flux (HF, green in figures) state. The HEASARC tool ADDSPEC is used to make a stacked spectrum for each of the three flux levels, along with corresponding background spectra and response matrix files. All the spectra are grouped to have a minimum count of 50 per bin.

\begin{figure}
\centering
\includegraphics[width=\hsize]{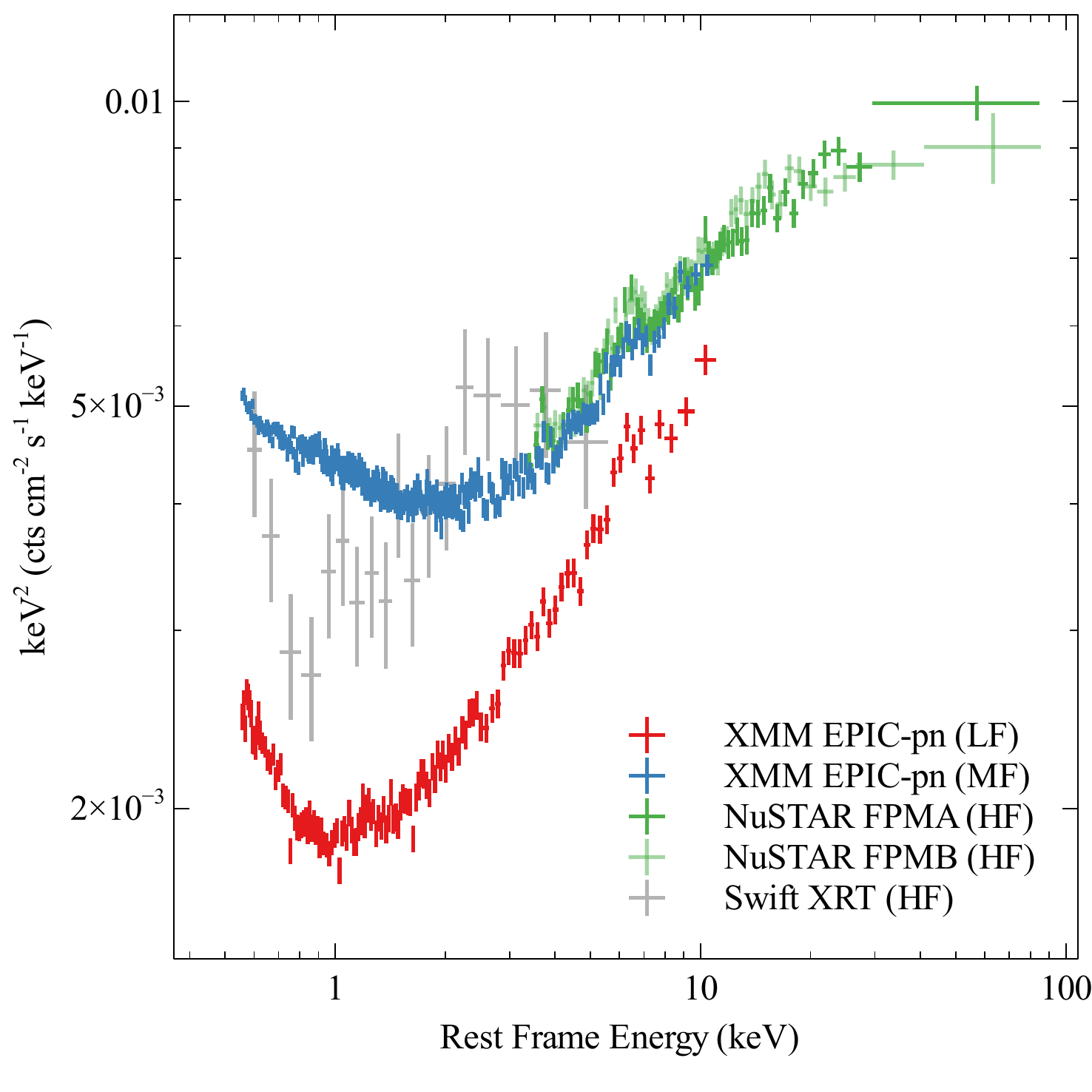}
\caption{The unfolded spectra of all the observations considered in this work against a constant model . Red: LF \xmm EPIC-pn spectrum; blue: MF \xmm EPIC-pn spectrum; green and light green: HF \nustar FPMA and FPMB spectra; grey: HF \swift XRT spectrum.}
\label{pic_eeuf}
\end{figure}

The spectra of all the observations considered in this work unfolded through a constant model are shown in Fig. \ref{pic_eeuf}. The soft band (e.g. 0.5--2~keV) shows a larger flux variability (2 times) than the iron band (1.4 times). The simultaneous HF \swift spectrum is shown in grey points in figures hereafter. The MF EPIC-pn spectrum and the HF FPM spectra show a similar continuum at the iron band with a small difference on the flux level. However the spectral shape below 2~keV is very different in MF and HF spectra. The HF \swift XRT spectrum shows a dip feature at 0.6--0.8~keV compared to the MF EPIC-pn spectrum. In this section, we initially focus on the spectral fitting of the broad band LF and HF state spectra - which represent the extremes of the observed spectral variability - to provide a model template for the subsequent multi-epoch spectral analysis.

XSPEC V12.10.0.C \citep{arnaud} is used for spectral analysis, and $\chi^2$ is considered in this work. The Galactic column density towards 1H0419$-$577 is fixed at $N_{\rm H}=1.34\times10^{20}$\,cm$^{-2}$ \citep{willingale13}. The photoionization cross section is from \citet{crosssec1} and He cross section is from \citet{crosssec2}. The solar abundances of \citet{wilms} were used. For local Galactic absorption, the \texttt{tbnew} model \citep{wilms} is used. An additional constant model \texttt{constant} has been applied to vary normalizations between the simultaneous spectra obtained by different instruments to account for calibration uncertainties. The following cosmology constants are considered: $H_o=73$\,km\,s$^{-1}$\,Mpc$^{-1}$, $\Omega_{\rm matter}=0.27$, and $\Omega_{\rm vacuum}=0.73$. Errors are calculated by estimating the 90\% confidence range of parameters using the ERROR command in XSPEC.

\subsection{\xmm Low Flux State Spectral Analysis} \label{lf_analysis}

\begin{figure*}
\centering
\includegraphics[width=\hsize]{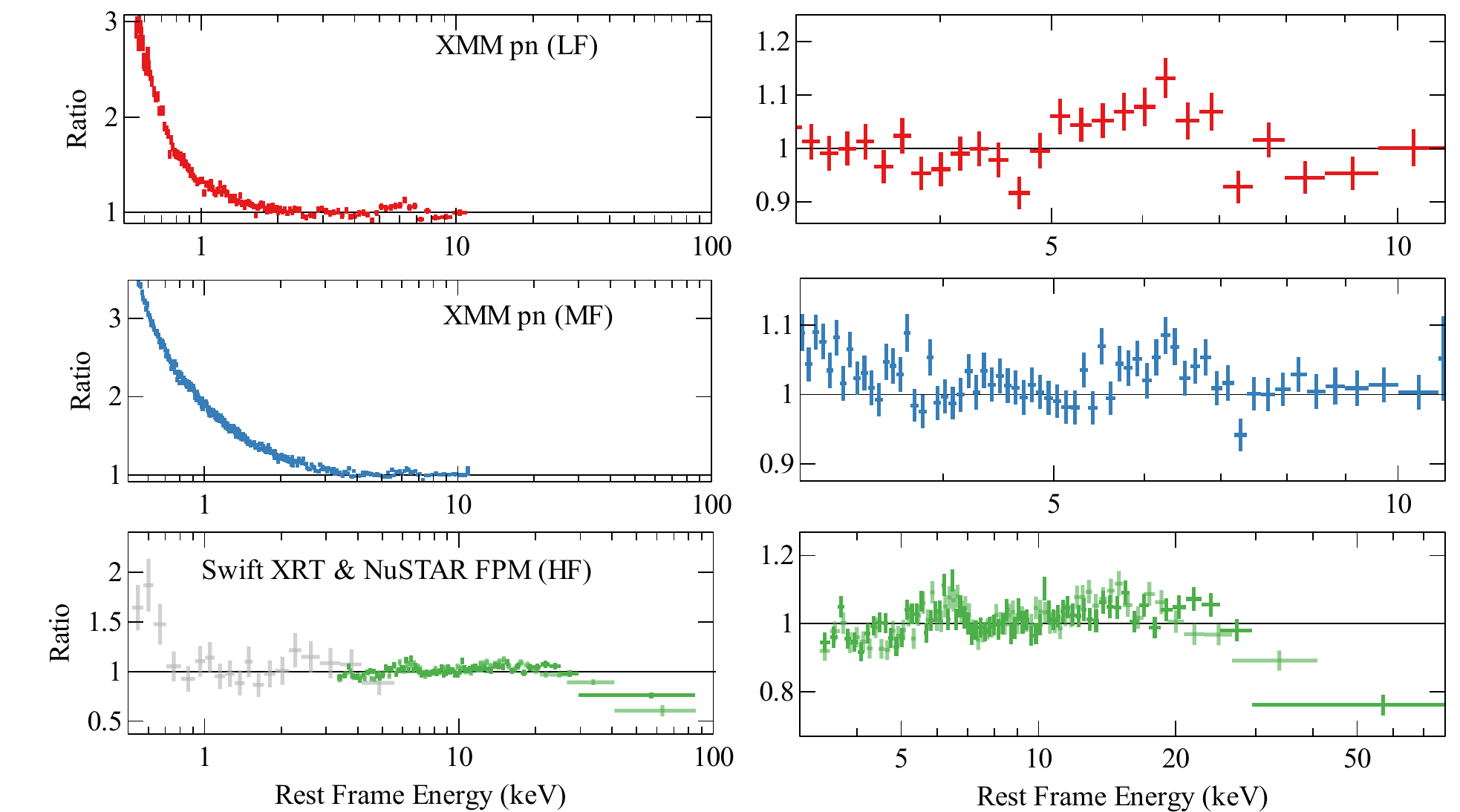}
\caption{The ratio plots of three flux state spectra against Galactic absorbed \green{power-law} models. All three sets of spectra show an iron emission line profile at the iron band and strong soft excess at the low energy band. The HF state spectra show a low high-energy turnover.}
\label{pic_pl}
\end{figure*}

Firstly, we fit the 2--10\,keV band with a Galactic absorbed \texttt{powerlaw} model and extend the ratio plot to 0.5\,keV without changing the fit. The ratio plot is shown in the top left panel of Fig. \ref{pic_pl}. It shows a very strong soft excess below 2~keV. The right panel shows the zoom-in of the iron band. A broad emission line is visible between 4 and 10\,keV. By fitting the emission feature with a simple gaussian line model \texttt{zgauss} with the rest frame energy of the Fe K$\alpha$ emission line ($E_{\rm line}=6.4$\,keV) and the redshift fixed at the source redshift, we obtained a best-fit line model with a line width of $\sigma=0.39^{+0.20}_{-0.12}$\,keV and an equivalent width of $130\pm10$\,eV. No additional narrow line component is required.

\begin{figure}
\centering
\includegraphics[width=\hsize]{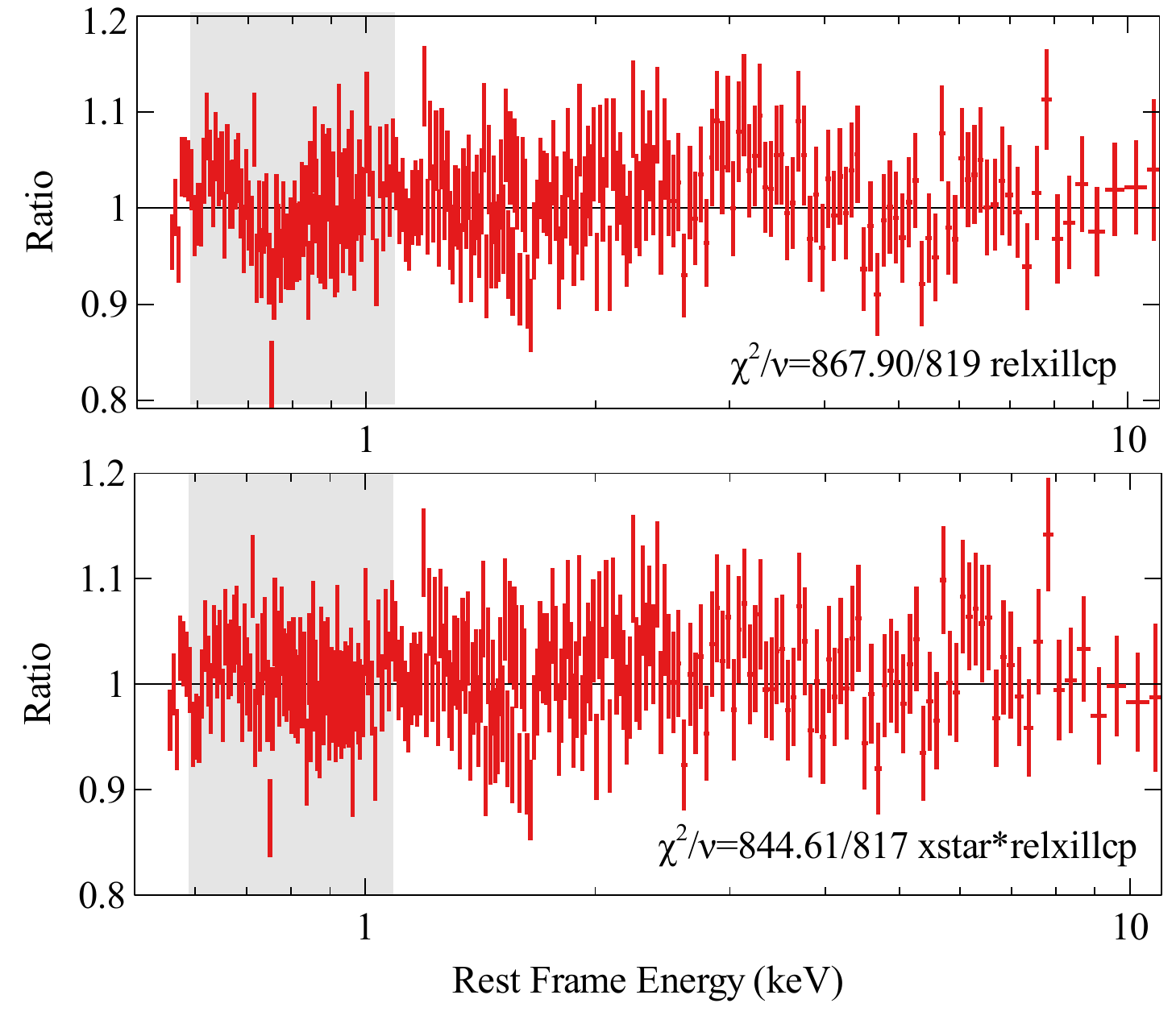}
\caption{The ratio plot of the LF spectrum against the best-fit \texttt{relxillcp} (top) and \texttt{xstar*relxillcp} (bottom) model. An additional warm absorber can improve the fit at the <1~keV band (grey shaded region).}
\label{pic_pn1_warmabs}
\end{figure}

Secondly, we fit the LF spectrum with the relativistic reflection model \texttt{relxillcp} (V1.0.4) \citep{dauser1,garcia14}, following the indications in \citet{fabian05} and \citet{walton13}. The relativistic reflection model \texttt{relxillcp} calculates the relativistic disc reflection spectrum given a thermally compotonized continuum, \texttt{nthcomp} \citep{zdziarski99,zycki99}. A temperature (low energy rollover) of $kT_0=0.05$~keV and a disc blackbody distribution is assumed for the seed photons. The relativistic effects, including both the Gravitational Redshift and the relativistic Doppler Effects, are all included in the \texttt{relxillcp} model \citep{dauser1}. We assume a simple broken power-law shaped emissivity profile for simplicity. The reflection fraction parameter is defined as the ratio of intrinsic intensity emitted from the corona towards the disc compared to the observer \citep{dauser16}. The coronal electron temperature $kT$ is fixed at a high value (100~keV) because of the lack of high-energy coverage during this epoch. The \texttt{relxillcp} model gives a good fit with $\chi^{2}/\nu$=887.90/819. The ratio plot against the best-fit relativistic reflection model shows a weak absorption feature between 0.6--0.9~keV. See the top panel of Fig. \ref{pic_pn1_warmabs}. 

Finally, by following the discovery of a weak and cool warm absorber by \citet{gesu13}, we also fit the absorption feature in the soft band with a warm absorber model. We constructed custom ionized absorption grids with \texttt{xstar} \citep{xstar}. The grids are calculated assuming solar abundances, a fixed turbulent velocity of 200~km\,$^{-1}$, and an ionzing luminosity of $10^{43}$\,erg\,s$^{-1}$. A power-law input spectrum with a photon index of $\Gamma=2$ is used. Free parameters are the column density $N_{\rm H}$ and $\log\xi^{\prime}$\footnote{The prime symbol is to distinguish the warm absorber ionization parameter from the disc ionization parameter $\log(\xi)$. The ionization $\xi$ is in unit of erg\,cm\,s$^{-1}$.}. This warm absorber improves the fit by $\Delta\chi^2=23$ with 2 more free parameters. See Fig.\,\ref{pic_pn1_warmabs}. An ionization state of $\log\xi^{\prime}=1.40^{+0.04}_{-0.13}$ is required for the warm absorber. The additional warm absorber fits the absorption feature with a series of low ionized absorption lines, such as O\textsc{iv-vi} lines and does not significantly change the key parameters of the continuum model. The best-fit parameters can be found in Table \ref{tab_fit_split} and the ratio plot can be found in the bottom panel of Fig.\,\ref{pic_pn1_warmabs}.

\begin{table}
\caption{The best-fit \texttt{xstar*relxillcp} parameters for the LF and HF spectra respectively. q1 and q2 are the inner and outer emissivity index respectively.}
\label{tab_fit_split}
\begin{tabular}{cccc}
\hline\hline
Model & Parameter  &  LF & HF \\
\hline
\texttt{xstar} & $N_{\rm H}$ ($10^{21}$\,cm$^{-2}$) & $1.5\pm0.3$ & <7.0 \\
& $\log(\xi^\prime$/erg cm s$^{-1}$) & $1.40^{+0.04}_{-0.13}$ & <1.8 \\
\hline
\texttt{relxillcp} & q1 & $>6$ & $>7.1$ \\
&q2 & $3.9^{+0.6}_{-0.7}$ & $3.3\pm0.4$  \\
&$R_{\rm break}$ (r$_{\rm g}$) & $2.1^{+1.0}_{-0.3}$ & $3.1\pm0.8$ \\
&a$_*$ & >0.98 & >0.96 \\
&$i$ (deg) & $22^{+7}_{-4}$ & $28^{+6}_{-9}$ \\
&$\Gamma$ & $2.06^{+0.02}_{-0.04}$ & $1.76\pm0.09$ \\
&$\log(\xi$/erg cm s$^{-1}$) & $0.9\pm0.5$ & $3.0^{+0.2}_{-0.3}$  \\
&$Z_{\rm Fe}$ ($Z_{\odot}$) & $0.8^{+0.7}_{-0.4}$ & $0.7^{+0.8}_{-0.2}$ \\
&$kT$ (keV) & 100 (fixed) & $27^{+43}_{-3}$  \\
&$f_{\rm refl}$ & $10^{+3}_{-2}$ & $3.2^{+0.3}_{-1.2}$ \\
\hline
&$\chi^{2}/\nu$ & 844.61/817 & 964.44/954\\
\hline\hline
\end{tabular}
\end{table}

An additional distant reflector \texttt{xillver} \citep{garcia13} was added for further test. All the parameters of the distant reflection component are linked to the corresponding parameters in the relativistic disc reflection component, except the normalization and the ionization parameter. An additional distant reflection component only improves the fit by $\Delta\chi^{2}=4$ with 2 more free parameters. The normalization of the \texttt{xillver} is $<1.0\times10^{-5}$ with 90\% confidence level. We conclude that no additional narrow reflection components are required in our analysis.

The current version of \texttt{relxillcp} assumes a constant electron density $n_{\rm e}=10^{15}$\,cm$^{-3}$ for the top layer of the BH accretion disc. However, recent spectral analysis of both Seyfert 1 AGNs with strong soft excess \citep[e.g. IRAS~13224$-$3809,][]{jiang18} and XRBs \citep[CygX-1 in the intermediate state,][]{tomsick18} show that the electron density assumed in the reflection model can have an important effect on some of the results obtained from spectral fitting. A more developed version \texttt{relxillD} \citep{garcia1} which allows the density to vary to between $n_{\rm e}=10^{15}$ and $10^{19}$\,cm$^{-3}$ is used to test any possible high electron density in 1H0419$-$577. A simple power-law shaped continuum is assumed for the coronal emission in \texttt{relxillD}. The same parameters are allowed to vary during the fit and we obtained $n_{\rm e}<10^{15.3}$ at a 90\% confidence level. The high density reflection model only improves the fit by $\Delta \chi^{2}=4$ with one more free parameter compared to \texttt{relxillcp}. We conclude that no higher electron density than $n_{\rm e}=10^{15}$\,cm$^{-3}$ is required for the spectral fitting. This result is appropriate for a disc around a BH of mass $>10^8$M$_{\odot}$, as in the case for 1H0419$-$577 -- a disc electron density of $n_{\rm e}<10^{16}$\,cm$^{-3}$ is expected at $r=20r_{g}$ away from a BH with $M_{\rm BH}>10^8$M$_{\odot}$ according to the solution by \citet{svensson94}. See Fig.\,1 in \citet{garcia1} for instance.

\begin{figure*}
\centering
\includegraphics[width=5.5cm]{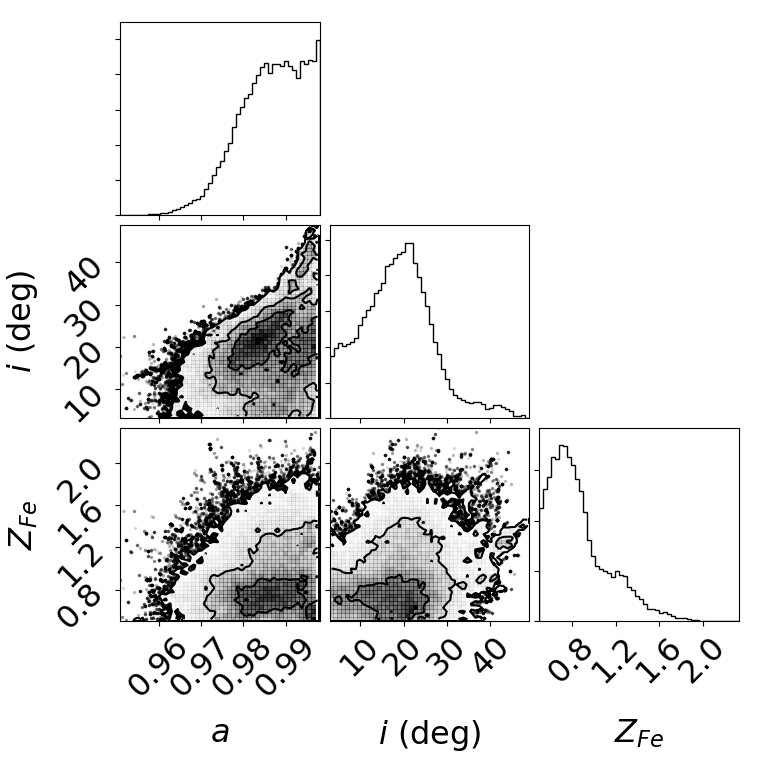}
\includegraphics[width=5.5cm]{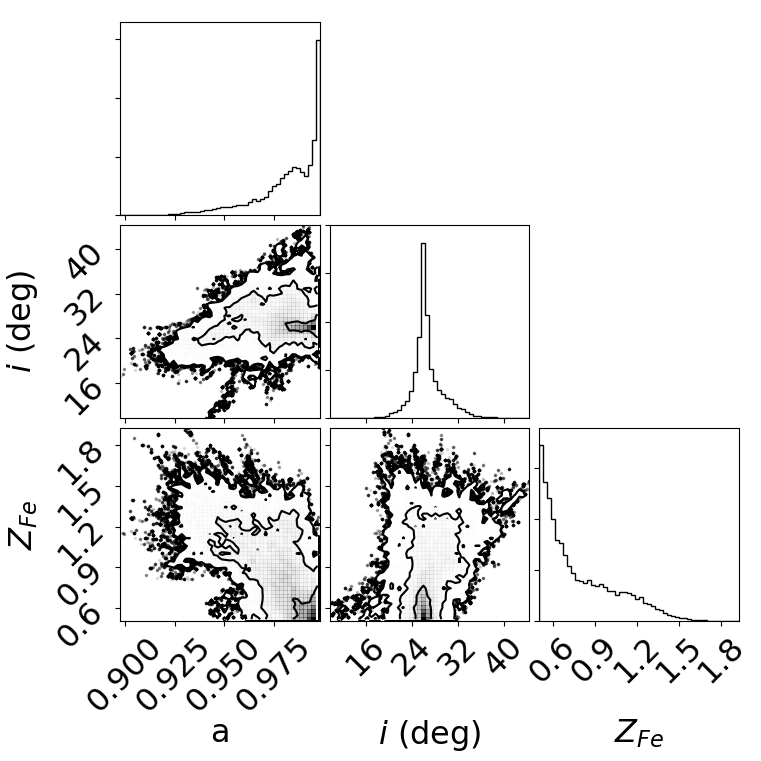}
\includegraphics[width=5.5cm]{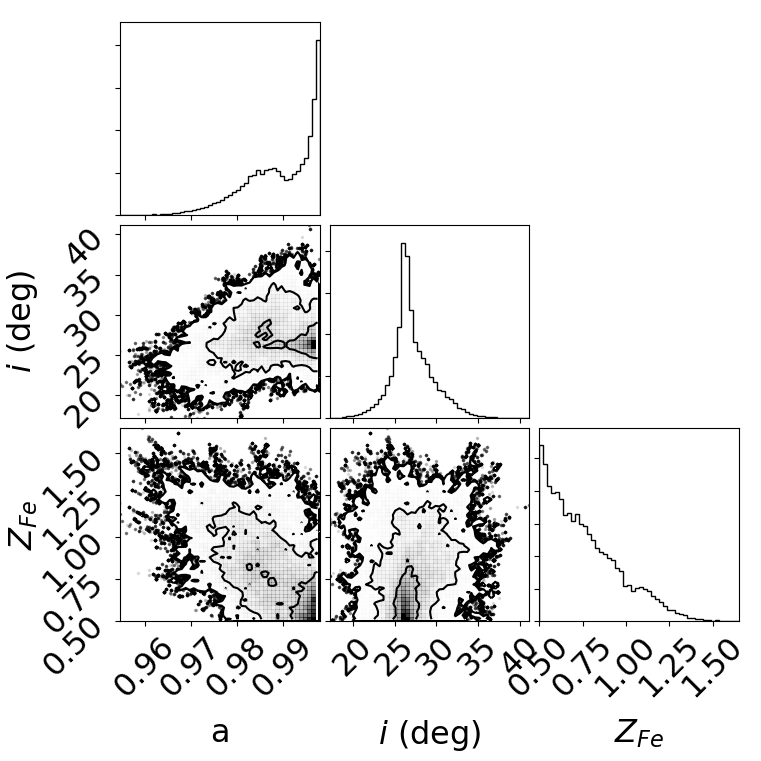}
\caption{Output distributions for the MCMC analysis of the best-fit models of the broad band spectra of 1H0419$-$577. Contours correspond to 1, 2 and 3$\sigma$. Only the spin $a_*$, the disc viewing angle $i$, and the disc iron abundance $Z_{\rm Fe}$ are shown here. Left: only the LF \xmm EPIC-pn spectrum; middle: only the HF \nustar and \swift spectra; right: multi-epoch spectral analysis of three flux state spectra.}
\label{pic_spin_incl}
\end{figure*}

We checked the constraints of all the parameters obtained in the LF spectral analysis by using the MCMC algorithm. The XSPEC/EMCEE code by Jeremy Sanders based on the python implementation \citep{foreman12} of the Goodman-Weare affine invariant MCMC ensemble sampler \citep{goodman10} was used for this purpose \footnote{The code can be found on following page. https://github.com/jeremysanders/xspec\_emcee}. We use 100 walkers with a length of 25000, burning the first 1000. A convergence test has been conducted and the Gelman-Rubin scale-reduction factor $R<1.3$ for every parameter. No obvious degeneracy was found. The contour plots of the two relativistic parameters $a_*$, $i$ and the disc iron abundance $Z_{\rm Fe}$ are shown in the left panel of Fig. \ref{pic_spin_incl}. By fitting only the EPIC-pn low flux state spectrum with \texttt{relxillcp}, we obtained $a_{*}>0.98$ and $i=22^{+7}_{-4}$ by running the ERROR command in XSPEC, which are consistent with our MCMC analysis.

\subsection{\nustar and \swift high flux state spectral analysis} \label{hf_analysis}

The \nustar and simultaneous \swift observations are taken when the source is in a high flux state (see Fig.\,\ref{pic_eeuf} for the unfolded spectra). We first fitted the \nustar FPM and \swift XRT spectra simultaneously with a Galactic absorbed power-law model. The ratio plot is shown in the bottom panel of the Fig.\,\ref{pic_pl}. A broad emission line feature is shown at the iron band with the low energy tail extending to 5~keV. By fitting the broad emission line feature with a simple \texttt{gauss} model, the fit is improved by $\Delta\chi^2=68$ with 3 more free parameters. The ratio plot against a power-law model also shows a Compton hump above 10~keV and a very low energy turn-over above 30~keV (see the bottom right panel for the zoom-in of the FPM spectra). 

Following the analysis of the LF state spectrum in Section \ref{lf_analysis}, we fitted the HF spectra with the same model as in Section \ref{lf_analysis}. An additional \texttt{constant} model is added in XSPEC to account for cross-calibration uncertainty. \texttt{xstar*rexillcp} offers a good fit with $\chi^2/\nu=964.44/954$. The warm absorber fits the absorption feature at 0.6--0.8~keV, similar to the warm absorber in the LF state spectrum. However, due to a low signal-to-noise of the XRT spectrum, we only obtained an upper limit on the column density and the ionization of the warm absorber. 

\begin{figure*}
\centering
\includegraphics[width=6cm]{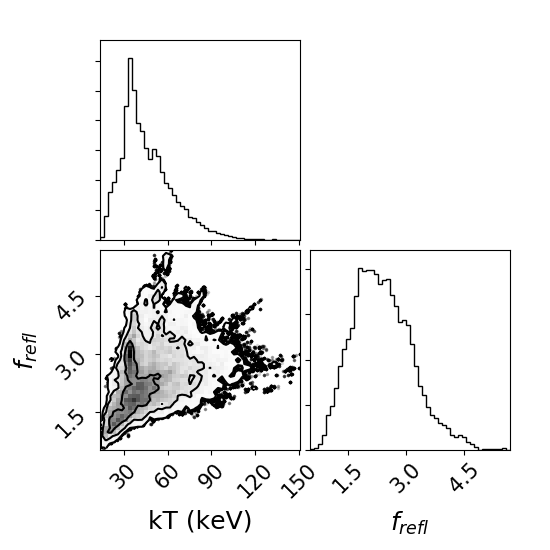}
\includegraphics[width=6cm]{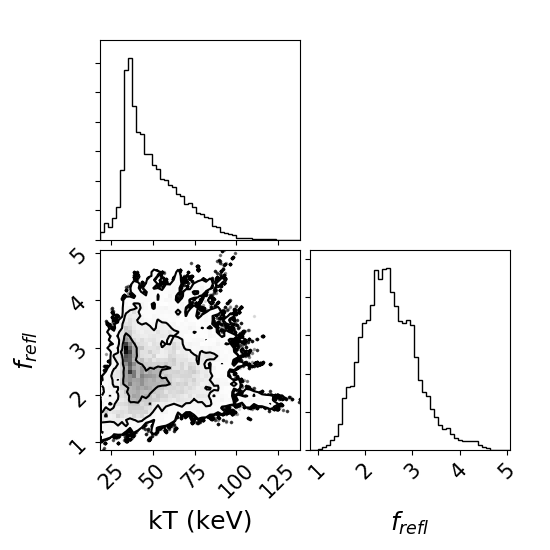}
\caption{Same as Fig. \ref{pic_spin_incl} but for the disc reflection fraction and energy cutoff parameters. Left: only the \nustar and \swift (HF) spectra; right: multi-epoch spectral analysis of all three flux state spectra.}
\label{pic_ecut_refl}
\end{figure*}

We conducted a similar MCMC analysis for this fit as in Section \ref{lf_analysis}. The constraints of the spin, the disc viewing angle, and the disc iron abundance are shown in the middle panel of Fig.\,\ref{pic_spin_incl}. The constraints on the three parameters are weaker compared to the fit of the LF spectrum but consistent within 1$\sigma$ uncertainty range. These quantities are not expected to vary on observable timescales, and so this gives us added confidence in our results. The low energy turn-over shown in the bottom panel of Fig. \ref{pic_pl} could be due to either a low energy cutoff in the coronal emission or the Compton hump in a reflection dominated spectrum. A contour plot on the coronal electron temperature $kT$ and the disc reflection fraction $f_{\rm refl}$ parameter plane is shown in the left panel of Fig.\,\ref{pic_ecut_refl}. Note that there is a weak degeneracy at 3$\sigma$ confidence level. However, the reflection fraction is constrained at a low value $f_{\rm refl}<3$ within 1$\sigma$ uncertainty range, precluding a reflection dominated scenario.

\subsection{Multi-epoch spectral analysis} \label{final}

\begin{table}
\caption{The same as Table \ref{tab_fit_split} but for the multi-epoch joint spectral fitting. The flux of the best-fit model is calculated by using \texttt{cflux} model in XSPEC at 1--10~keV band in erg cm$^{-2}$ s$^{-1}$.}
\label{tab_fit}
\begin{tabular}{c | c c | c }
\hline\hline
Parameter  &  LF & MF & HF \\
\hline
 $N_{\rm H}$ ($10^{21}$\,cm$^{-2}$) & $1.1\pm0.2$ & <0.15 & <1.3 \\
$\log(\xi^\prime$/erg cm s$^{-1}$) & $1.40^{+0.04}_{-0.13}$ & <1.8 & <1.8 \\
\hline
q1 & $5.7^{+2.3}_{-0.5}$ & $5\pm2$ & $7.8^{+1.2}_{-0.3}$\\
q2 & $2.7^{+0.2}_{-0.3}$ & $4.6^{+0.5}_{-0.2}$ & $3.1^{+0.2}_{-0.4}$ \\
$R_{\rm break}$ (r$_{\rm g}$) & $5.5^{+0.2}_{-1.2}$ & <12 & $4.2^{+0.2}_{-1.8}$\\
a$_*$ & & >0.987 & \\
$i$ (deg) & & $26^{+8}_{-4}$ & \\
$\Gamma$ & $2.07^{+0.02}_{-0.05}$ & $2.308^{+0.008}_{-0.011}$ & $1.887^{+0.063}_{-0.007}$ \\
$\log(\xi$/erg cm s$^{-1}$) & $1.04^{+0.07}_{-0.11}$ & <0.03 & $2.85^{+0.03}_{-0.15}$  \\
$Z_{\rm Fe}$ ($Z_{\odot}$) & & $0.7^{+0.5}_{-0.3}$ & \\
$kT$ (keV) & & $30^{+22}_{-7}$ & \\
$f_{\rm refl}$ & $10^{+4}_{-2}$ & $5.3^{+1.2}_{-0.2}$ & $2.8^{+1.0}_{-1.3}$ \\
$\log(F_{\rm 1-10~keV})$  & $-10.914^{+0.003}_{-0.002}$ & $-10.740^{+0.001}_{-0.002}$ & $-10.70\pm0.01$\\
\hline
$\chi^{2}/\nu$ & & 3337.79/3272\\
\hline\hline
\end{tabular}
\end{table}

\begin{figure*}
\centering
\includegraphics[width=\hsize]{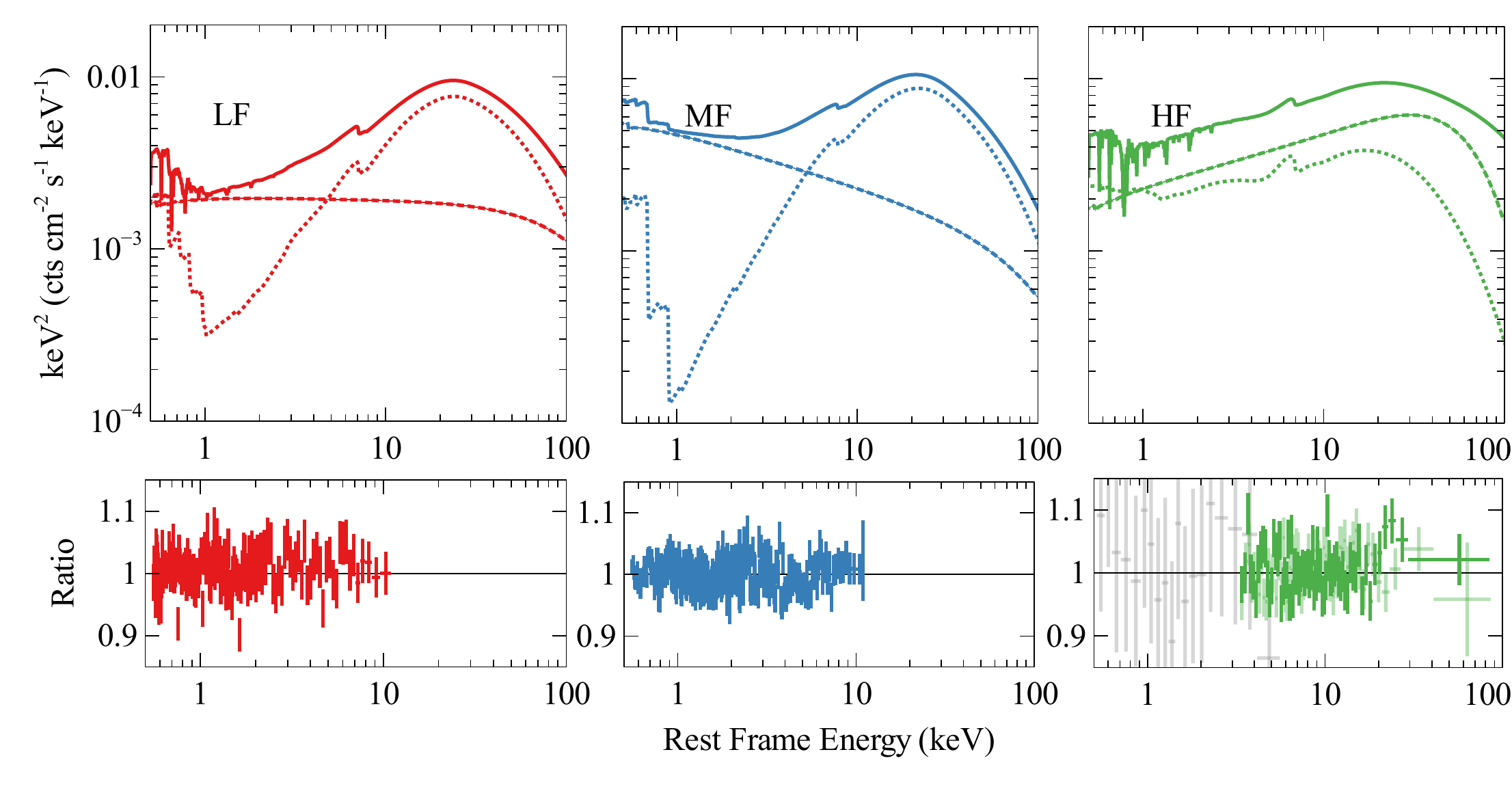}
\caption{Top: the best-fit \texttt{xstar*relxillcp} model for three sets of spectra (red: LF; blue: MF; green: HF) obtained in the multi-epoch spectral analysis. The dashed lines show the best-fit unabsorbed coronal emission model \texttt{nthcomp} modelled in \texttt{relxillcp}; the dotted lines show the best-fit unabsorbed disc reflection component. Bottom: three ratio plots of three flux state spectra against the best-fit \texttt{xstar*relxillcp} models correspondingly.}
\label{pic_ref}
\end{figure*}

\begin{figure}
\centering
\includegraphics[width=\hsize]{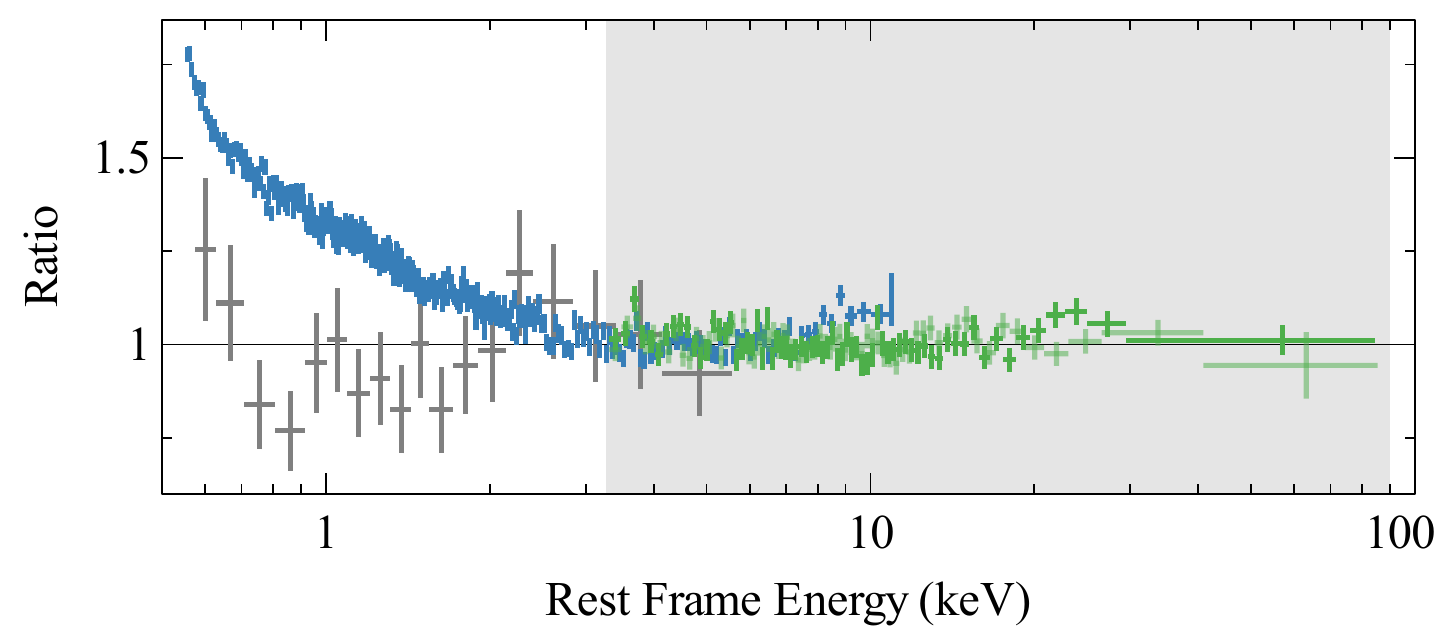}
\caption{The ratio plot of MF and HF spectra against the best-fit \texttt{relxillcp} model obtained by fitting the MF EPIC-pn (blue) and HF FPM spectra (green) simultaneously at the 3--78~keV band (grey shaded region). The plot is extended to 0.5\,keV without changing the model. See text for more details.}
\label{pic_test}
\end{figure}

\begin{figure}
\centering
\includegraphics[width=\hsize]{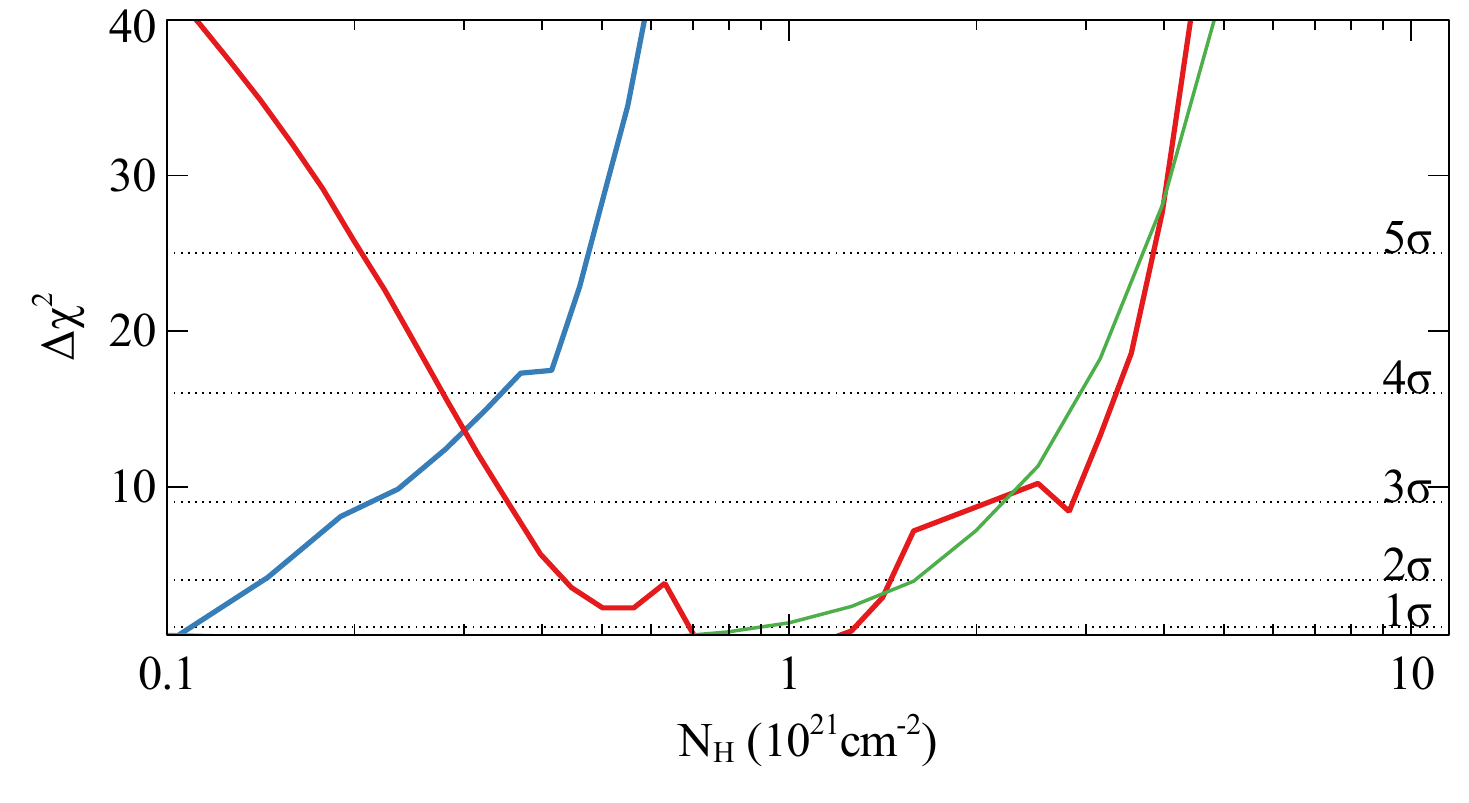}
\caption{The $\chi^2$ contour plot for the column density of the warm absorber in the multi-epoch joint spectral analysis (red: LF; blue: MF; green: HF). The 1--5$\sigma$ measurement rage is marked in dotted lines.}
\label{pic_nh}
\end{figure}

In previous sections, we have obtained a good fit for 1H0419$-$577 LF and HF state spectra by using the combination of a warm absorber and a relativistic disc reflection model \texttt{xstar*relxillcp}. The analysis has found consistent results of the key reflection parameters (spin, disc viewing angle, iron abundance). In this section we therefore undertake the multi-epoch spectral analysis of all three flux state spectra to more robustly probe the spectral variability shown by this source.

The same disc reflection model \texttt{relxillcp} was used. The BH spin parameter, the disc viewing angle, and the disc iron abundance are not expected to vary in the time scale of our observations and they are linked during the joint spectral analysis. The LF and MF state \xmm observations were taken with lack of simultaneous high energy observations above 10~keV. We therefore also linked the coronal electron temperature parameter $kT$ for all three flux state spectra. The best-fit continuum model parameters for the lower flux state spectra obtained after linking the coronal temperature parameters are consistent with the values obtained by fitting them alone within their 90\% confidence errors. The other parameters are all allowed to vary during the fit. The best-fit model parameter values are shown in Table \ref{tab_fit}. The best-fit \texttt{xstar*relxillcp} models and the ratio plot against the best-fit models are shown in Fig.\,\ref{pic_ref}. A similar MCMC analysis has been conducted for the multi-epoch spectral analysis as in previous sections. 200 walkers are used for a larger number of degrees of freedom to keep the Gelman-Rubin scale-reduction factor $R<1.3$ in the convergence test. The constraints on the relativistic parameters and $Z_{\rm Fe}$, and a $kT$ and $f_{\rm refl}$ are shown in the right panels of Fig.\,\ref{pic_spin_incl} and Fig.\,\ref{pic_ecut_refl} respectively. The parameter measurements given by running ERROR command in 90\% confidence level are all consistent with MCMC analysis results.

By combining all three flux state spectra, we obtained a close-to-maximum black hole spin of $a_*>0.987$ and a disc viewing angle of $i=26^{+8}_{-4}$$^{\circ}$. By fitting the \suzaku observation with a similar model, \citet{walton13} obtained a viewing angle of $\approx45$$^{\circ}$ and a weaker spin constraint $a_{*}>0.89$. The difference between the two inclination angles could be due to the model development or instrumental systematic uncertainty (see discussion in \citealt{brenneman13}) or stacking two different observations at different flux and ionization states.

A very high reflection fraction $f_{\rm refl}\approx10$ is obtained for the LF state while a lower reflection fraction $f_{\rm refl}\approx3-6$ is measured for the HF and MF state. The light-bending model can potentially explain the higher reflection fraction found in the LF state of 1H0419$-$577. In the light-bending model, more primary continuum photons will be lost to the BH and the trajectories of photons will be bent towards the central object when the corona is closer to the BH \citep[e.g. the LF state in 1H0419$-$577 corresponds to the reflection dominated regime I or I/II in][]{miniutti04}. In order to rule out the possibility that the $f_{\rm refl}$ obtained above is only due to the variable soft excess, we fitted all three spectra at the 3--10~keV band with a simple \texttt{tbabs*(powerlaw+zgauss)} model, where \texttt{zgauss} accounts for the broad Fe K$\alpha$ emission line at the iron band. The redshift $z$ of the \texttt{zgauss} model is fixed at the source redshift and the rest-frame energy is fixed at 6.4~keV. The equivalent widths of the best-fit \texttt{zgauss} against the simple Galactic power-law continuum are $130\pm10$~eV for the LF state, $62^{+23}_{-41}$~eV for the MF state, and $78^{+23}_{-21}$~eV for the HF state. A higher equivalent width of the Fe K$\alpha$ emission line at the LF state matches the result obtained by fitting the broad band spectra.

A thin and low ionisation warm absorber with a column density of $N_{\rm H}\approx10^{21}$\,cm$^{-2}$ and an ionization state of $\log(\xi)\approx1.5$ is required to fit the dip features in the LF state spectra. A $\chi^2$ contour plot for the warm absorber column density is shown in Fig.\,\ref{pic_nh}. We can conclude that no warm absorber or at least an even smaller column ($N_{\rm H}\approx10^{20}$\,cm$^{-2}$) is required for the MF state. The result matches the RGS spectral analysis of the MF observations ($N_{\rm H}\approx10^{19.9}$\,cm$^{-2}$) in \citet[][]{gesu13}. The warm absorber in the HF state can account for the spectral difference between the MF and HF spectra at the 0.6--0.8~keV band, in addition to the continuum variability.

\subsection{Further comparison between the MF and HF spectra} \label{comparison}

The HF FPM and MF EPIC-pn spectra show a similar spectral shape with only 15\% different flux level in the 3--10~keV band, although two observations were taken 5 years apart. However in the soft band (<2keV), the HF XRT and the MF EPIC-pn observations show a very different spectral shape. See Fig.\,\ref{pic_eeuf} for the unfolded spectra. By modelling the spectra with \texttt{warmabs*relxillcp}, we found that both the disc reflection component and the warm absorber need to be variable to account for the large spectral variability below 2~keV despite a similar spectral shape at the iron band. For example, different photon index $\Gamma$, different column of the warm absorber $N_{\rm H}$, and different disc ionization state $\xi$. are all required in the multi-epoch spectral fitting.

However there might be other solutions to the soft band spectral variability. For instance, only a different warm absorber can account for the soft band spectral variability with a similar disc reflection and coronal emission for both two flux states. In order to test this scenario, we first ignored the HF XRT spectrum, and fitted the MF EPIC-pn spectrum (3--10~keV) and the HF FPM spectrum (3--78~keV) simultaneously with the same \texttt{relxillcp} model. An additional \texttt{constant} model in XSPEC is used to account for the flux difference between two epochs. We obtained a good fit with $\chi^{2}_{\rm red}=1.10$. The best-fit \texttt{relxillcp} requires a photon index of $\Gamma\approx1.87$, an ionization of $\log(\xi)\approx2.7$, a reflection fraction of $f_{\rm refl}\approx3.3$, a black hole spin of $a_*>0.96$, and a disc viewing angle of $i\approx28$$^{\circ}$. The best-fit model is similar with the best-fit continuum model obtained by using \texttt{warmabs*relxillcp} in Section \ref{hf_analysis}. The ratio plot extended to 0.5\,keV without changing the fit is shown in Fig.\,\ref{pic_test}. The HF XRT spectrum is added for reference in the figure. First, we notice that the best-fit \texttt{relxillcp} fails to fit the hard band of the MF EPIC-pn spectrum with 20\% residuals above 7\,keV. It indicates a different emissivity profile is required for the relativistic disc line modelling. Second, by extending the ratio plot to 0.5\,keV without changing the fit, the best-fit \texttt{relxillcp} model obtained by fitting the hard band fails to fit the soft excess of the MF EPIC-pn spectrum but more agrees with the HF XRT spectrum. The fit of the HF XRT spectrum can be improved by fitting the remaining residuals shown in Fig.\,\ref{pic_test} with an additional thin warm absorber (see Section \ref{hf_analysis}). We therefore conclude that both a variable continuum model \texttt{relxillcp} and and a variable thin warm absorber \texttt{warmabs} are required to account for the spectral variability below 2~keV.

\section{Discussion and conclusions}

We fitted all three flux state spectra of the Seyfert 1 1H0419$-$577 successfully with a combination of a thin warm absorber and a variable relativistic disc reflection model. In this section, we discuss the accretion rate of the disc, the black hole spin obtained by fitting the multi-epoch broad band spectra with relativistic disc reflection model, the broad band spectral variability, and the properties of the cool corona region. 

\subsection{Eddington ratio estimation}

We calculated the Eddington ratio $\lambda_{\rm Edd}$ by applying an averaged bolometric luminosity correction factor $\kappa=20$ \citep{bolometric} to the 2--10~keV band absorption corrected luminosity $2.45\sim3.84\times10^{44}$\,erg\,s$^{-1}$. A black hole mass of $M_{\rm BH}=1.3\times10^{8}M_{\odot}$ \citep{grupe10} is considered. We obtained $\lambda_{\rm Edd} = \kappa L_{\rm x}/L_{\rm Edd}\approx20\times0.015\sim0.024=0.30\sim0.48$. Note that the bolometric luminosity correction factor $\kappa$ can be even higher than 20 when the $\lambda_{\rm Edd}$ is at a high value. We therefore conclude that the disc around the SMBH in 1H0419$-$577 is accreting at an accretion rate approaching the Eddington limit.  

\subsection{Black hole spin measurement}

Previously, \citet{walton13} obtained a BH spin of $a>0.88$ by fitting the \suzaku observation of 1H0419$-$577 at high flux state with relativistic disc reflection model \texttt{reflionx} and \citet{fabian05} obtained an inner accretion disc radius of $R_{\rm in}< 2 R_{\rm g}$ by fitting the low flux state spectra captured by \xmm with an ionized reflection model convolved with \texttt{kdblur} and concluded a spin of $a>0.95$. In this work, we have conducted a robust measurement of the central BH spin by conducting careful MCMC analysis and obtained a black hole spin of $a_*>0.987$ (see the right panel of Fig.\,\ref{pic_spin_incl} for MCMC analysis). By excluding the spectra below 3~keV where the strong soft excess is, we still obtained a high black hole spin of $a_*>0.96$ (see Section \ref{comparison}).

1H0419$-$577 is a Seyfert 1 galaxy hosting a SMBH with $M_{\rm BH}>10^8M_{\odot}$ \citep{grupe10}. A significant fraction of the SMBH spin measured by using relativistic reflection spectroscopy are very close to the maximum \citep[e.g. see the sample list in][]{brenneman13, walton13}. By compiling the AGN spin measurements obtained through reflection spectroscopy in the literature, \citet{reynolds14} pointed out that there is tentative evidence that the most massive black holes ($M_{\rm BH}>10^8M_{\odot}$) and the least massive black holes ($M_{\rm BH}<10^6M_{\odot}$) may have more modest spins. One of the possible explanations is the effect of host galaxy properties on the evolution of the black hole spin \citep{senana14}. However, 1H0419$-$577 shows both a high black hole spin and a large black hole mass ($\approx 10^{8}M_{\odot}$).

\subsection{The spectral variability}

In previous section, we introduce the spectral analysis of 3 different flux states captured by \textit{XMM-Newton}, \swift and \nustar observations. The soft band (0.5--2~keV) of 1H0419$-$577 shows a larger flux variability (2 times) than the iron band (1.4 times). See Fig.\, \ref{pic_eeuf} for unfolded spectra after correcting for the effective area of the detectors. The MF state and HF state shows a similar spectral slope at 1--10~keV band but a very different soft band. By fitting the broad K$\alpha$ emission line and the soft excess simultaneously with only one \texttt{relxillcp} model, we obtained a higher reflection fraction for the LF state compared to the higher flux states. The best-fit reflection fraction for the MF state spectrum is slightly higher than HF state. The equivalent width of the Fe K$\alpha$ also shows a similar correlation with the source flux level. Such an anti-correlation between the X-ray band flux and the disc reflection fraction can be explained by the light-bending effect in the vicinity of the black hole. The light-bending model has been discussed in previous literatures. For example, the LF state of 1H0419$-$577 corresponds to the regime I and I/II discussed in \citep{miniutti04}, where the spectrum is dominated by the disk reflection component. When the coronal region is closer to the central black hole, more continuum photons from the coronal region are lost to the event horizon. Both the direct emission from the coronal component and the reflection component decrease with decreasing flux while the reflection fraction increases. The light-bending effects have successfully explained the X-ray spectral variability in others sources as well, such as Mrk335 \citep{parker15} and IRAS13224$-$3809 \citep{chiang, jiang18}.     

The coronal emission shows an interesting variability versus flux in different epochs compared with other typical Seyfert 1 sources: the spectrum is hardest at the HF flux state ($\Gamma\approx1.8$) compared to the lower flux states ($\Gamma\approx2.0-2.3$), although the MF state spectrum has a slightly softer continnum than the LF state spectrum. It agrees with previous analysis of the same source: \citet{walton13} stacked the two archival \suzaku observations, one of which was taken at a 16\% higher flux state than the \nustar observation in this work and the other was at a 7\% lower flux state, and obtained a hard continuum too ($\Gamma=1.98$). \citet{fabian05} analyzed the first orbits of the \xmm observations, which were taken at this source's lowest flux state, and obtained a very soft continuum ($\Gamma\approx2.2$). The continuum emissions are commonly found to be softer at higher luminosities \citep[e.g.][]{shemmer06}. For example, the narrow-line Seyfert 1 IRAS~13224$-$3809 accretes at an accretion rate around the Eddington limit and shows a softer continuum at higher X-ray luminosities \citep{jiang18}. One explanation for an exception, as in 1H0419$-$577, is a possible advective flow in the innermost region \citep{esin97,narayan05}. The Compton parameter in such a flow increases as the accretion rate increases and thus produces a harder continuum. This scenario applies to either XRBs in the intermediate hard state or AGNs with a low accretion rate. Indeed the harder-when-brighter continuum is commonly seen in low-luminosity AGNs \citep[e.g. $L_{\rm x}/L_{\rm Edd}<10^{-2}$][]{connolly16}. However 1H0419$-$577 has a very high accretion rate. The detection of a broad iron K$\alpha$ emission line, a strong soft excess and high UV flux in its broad band SED \citep{turner18} indicate the existence of an inner disc. A second explanation is potential jet contribution to the X-ray spectrum \citep[e.g.][]{krawczynski04,zhang06}. However 1H0419$-$577 is a radio-quiet source with no significant detection of a radio jet. Additional observations covering more flux states of this source will be required to confirm and study the origin of the harder-when-brighter continuum. 

\subsection{Low high-energy cutoff}

\begin{figure}
\centering
\includegraphics[width=\hsize]{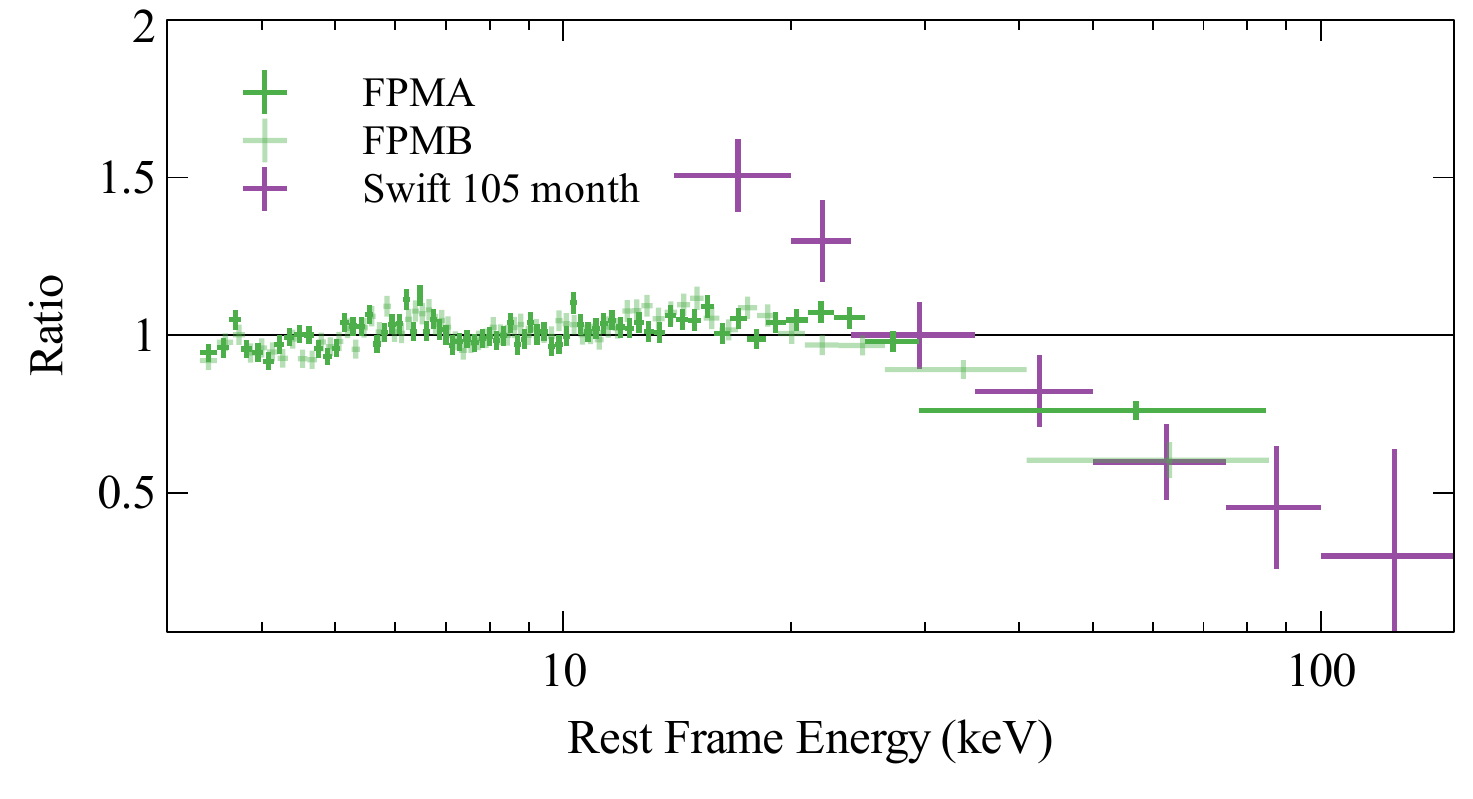}
\caption{The ratio plots of \nustar FPM spectra and \swift BAT 105-month survey spectrum (15--150~keV) against the best-fit galactic absorbed power-law model. The cross-calibration constant for BAT is 0.75.}
\label{pic_bat}
\end{figure}

In this work, we fitted the broad band spectra carefully with the relativistic reflection model \texttt{relxillcp} and tested any possible degeneracy between the energy cutoff and the disc reflection fraction by conducting an MCMC analysis. A low energy spectral turn-over could be due to either a cool coronal temperature or a reflection dominated spectrum. We conclude that the HF \nustar observation shows a cool corona with a temperature of $kT=30^{+22}_{-7}$\,keV and a disc reflection fraction of $f_{\rm refl}=2.8^{+1.0}_{-1.3}$, precluding the very high reflection scenario. 

In order to seek for any possible variability of the energy cutoff $E_{\rm cut}$ in a long timescale, we fitted the \swift 105-month BAT spectrum \citep{oh18} together with the \nustar spectrum. The ratio plot is shown in Fig.\,\ref{pic_bat}. The \swift BAT spectrum shows a very steep soft spectral shape and agrees with the \nustar FPM spectra above 30~keV. More \nustar observations are required to confirm any variability of $E_{\rm cut}$ in a wider flux range \cite[e.g.][]{fabian17}.

Such a low energy cutoff is also seen in other sources, such as Ark564 \citep[$kT=15\pm2$\,keV]{kara17}, GRS 1734$-$292 \citep[$kT=11.9^{+1.2}_{-0.9}$\,keV]{tortosa17}, IRAS~13197$-$1627 \citep[$kT<42$~keV][]{walton18b}, and 4C~50.55 \citep[$kT\approx30$\,keV]{tazaki10}. Except the Seyfert 1 galaxy GRS 1734$-$292 ($\lambda_{\rm Edd}\approx0.03$) and the Seyfert 1.8 Galaxy IRAS~13197$-$1627 ($\lambda_{\rm Edd}\approx0.05-0.1$), all the other sources mentioned above are accreting at a very high accretion rate. A very extreme example is the narrow line Seyfert 1 galaxy Ark564 which accretes at the Eddington limit and shows the coolest coronal temperature so far. One possibility is that a high accretion rate disc is cooling down the coronal region more than a low accretion rate disc by providing more seed photons. 

\subsection{Future observations}

In this work, we successfully explained the spectral variability, especially the soft band, with the combination of a variable disc reflection model and a thin warm absorber. The strong soft excess and soft band variability make this source a promising candidate with detections of disc reverberation lags, as seen in other sources, such as 1H0707$-$495 \citep[e.g.][]{fabian13,kara13} and Ark564 \citep[e.g.][]{kara17}. However a large black hole mass $>10^8M_{\odot}$ means a much longer observation is needed to detect any reverberation lag compared with the narrow-line Seyfert 1 galaxies mentioned above. Moreover, the hard band \nustar spectrum of 1H0419$-$577 reveals a cool coronal region. More \nustar observations at different flux states are required to monitor possible variation on the coronal temperature.

\section*{Acknowledgements}

J.J. acknowledges support by the Cambridge Trust and the Chinese Scholarship Council Joint Scholarship Programme (201604100032). D.J.W. acknowledges support from an STFC Ernest Rutherford Fellowship. A.C.F. acknowledges support by the ERC Advanced Grant 340442. M.L.P. acknowledges support by an ESA Research Fellowship. This work made use of data from the \nustar mission, a project led by the California Institute of Technology, managed by the Jet Propulsion Laboratory, and funded by NASA. This research has made use of the \nustar Data Analysis Software (NuSTARDAS) jointly developed by the ASI Science Data Center and the California Institute of Technology. This work made use of
data supplied by the UK Swift Science Data Centre at the University of Leicester.




\bibliographystyle{mnras}
\bibliography{1h0149.bib} 

\begin{thebibliography}{}
\makeatletter
\relax
\def\mn@urlcharsother{\let\do\@makeother \do\$\do\&\do\#\do\^\do\_\do\%\do\~}
\def\mn@doi{\begingroup\mn@urlcharsother \@ifnextchar [ {\mn@doi@}
  {\mn@doi@[]}}
\def\mn@doi@[#1]#2{\def\@tempa{#1}\ifx\@tempa\@empty \href
  {http://dx.doi.org/#2} {doi:#2}\else \href {http://dx.doi.org/#2} {#1}\fi
  \endgroup}
\def\mn@eprint#1#2{\mn@eprint@#1:#2::\@nil}
\def\mn@eprint@arXiv#1{\href {http://arxiv.org/abs/#1} {{\tt arXiv:#1}}}
\def\mn@eprint@dblp#1{\href {http://dblp.uni-trier.de/rec/bibtex/#1.xml}
  {dblp:#1}}
\def\mn@eprint@#1:#2:#3:#4\@nil{\def\@tempa {#1}\def\@tempb {#2}\def\@tempc
  {#3}\ifx \@tempc \@empty \let \@tempc \@tempb \let \@tempb \@tempa \fi \ifx
  \@tempb \@empty \def\@tempb {arXiv}\fi \@ifundefined
  {mn@eprint@\@tempb}{\@tempb:\@tempc}{\expandafter \expandafter \csname
  mn@eprint@\@tempb\endcsname \expandafter{\@tempc}}}

\bibitem[\protect\citeauthoryear{{Arnaud}}{{Arnaud}}{1996}]{arnaud}
{Arnaud} K.~A.,  1996, {XSPEC: The First Ten Years}

\bibitem[\protect\citeauthoryear{{Balucinska-Church} \&
  {McCammon}}{{Balucinska-Church} \& {McCammon}}{1992}]{crosssec1}
{Balucinska-Church} M.,  {McCammon} D.,  1992, \mn@doi [\apj] {10.1086/172032},
  \href {http://adsabs.harvard.edu/abs/1992ApJ...400..699B} {400, 699}

\bibitem[\protect\citeauthoryear{{Bambi} et~al.,}{{Bambi}
  et~al.}{2018}]{bambi18}
{Bambi} C.,  et~al., 2018, \mn@doi [Universe] {10.3390/universe4070079}, \href
  {http://adsabs.harvard.edu/abs/2018Univ....4...79B} {4, 79}

\bibitem[\protect\citeauthoryear{{Bardeen}, {Press}  \& {Teukolsky}}{{Bardeen}
  et~al.}{1972}]{bardeen72}
{Bardeen} J.~M.,  {Press} W.~H.,   {Teukolsky} S.~A.,  1972, \mn@doi [\apj]
  {10.1086/151796}, \href {http://adsabs.harvard.edu/abs/1972ApJ...178..347B}
  {178, 347}

\bibitem[\protect\citeauthoryear{{Blandford} \& {Znajek}}{{Blandford} \&
  {Znajek}}{1977}]{blandford77}
{Blandford} R.~D.,  {Znajek} R.~L.,  1977, \mn@doi [\mnras]
  {10.1093/mnras/179.3.433}, \href
  {http://adsabs.harvard.edu/abs/1977MNRAS.179..433B} {179, 433}

\bibitem[\protect\citeauthoryear{{Brenneman}}{{Brenneman}}{2013}]{brenneman13}
{Brenneman} L.,  2013, {Measuring the Angular Momentum of Supermassive Black
  Holes}, \mn@doi{10.1007/978-1-4614-7771-6.
}

\bibitem[\protect\citeauthoryear{{Brenneman} et~al.,}{{Brenneman}
  et~al.}{2011}]{brenneman11}
{Brenneman} L.~W.,  et~al., 2011, \mn@doi [\apj] {10.1088/0004-637X/736/2/103},
  \href {http://adsabs.harvard.edu/abs/2011ApJ...736..103B} {736, 103}

\bibitem[\protect\citeauthoryear{{Chiang}, {Walton}, {Fabian}, {Wilkins}  \&
  {Gallo}}{{Chiang} et~al.}{2015}]{chiang}
{Chiang} C.-Y.,  {Walton} D.~J.,  {Fabian} A.~C.,  {Wilkins} D.~R.,   {Gallo}
  L.~C.,  2015, \mn@doi [\mnras] {10.1093/mnras/stu2087}, \href
  {http://adsabs.harvard.edu/abs/2015MNRAS.446..759C} {446, 759}

\bibitem[\protect\citeauthoryear{{Connolly}, {McHardy}, {Skipper}  \&
  {Emmanoulopoulos}}{{Connolly} et~al.}{2016}]{connolly16}
{Connolly} S.~D.,  {McHardy} I.~M.,  {Skipper} C.~J.,   {Emmanoulopoulos} D.,
  2016, \mn@doi [\mnras] {10.1093/mnras/stw878}, \href
  {http://adsabs.harvard.edu/abs/2016MNRAS.459.3963C} {459, 3963}

\bibitem[\protect\citeauthoryear{{Dauser}, {Garcia}, {Wilms}, {B{\"o}ck},
  {Brenneman}, {Falanga}, {Fukumura}  \& {Reynolds}}{{Dauser}
  et~al.}{2013}]{dauser1}
{Dauser} T.,  {Garcia} J.,  {Wilms} J.,  {B{\"o}ck} M.,  {Brenneman} L.~W.,
  {Falanga} M.,  {Fukumura} K.,   {Reynolds} C.~S.,  2013, \mn@doi [\mnras]
  {10.1093/mnras/sts710}, \href
  {http://adsabs.harvard.edu/abs/2013MNRAS.430.1694D} {430, 1694}

\bibitem[\protect\citeauthoryear{{Dauser}, {Garc{\'{\i}}a}, {Walton},
  {Eikmann}, {Kallman}, {McClintock}  \& {Wilms}}{{Dauser}
  et~al.}{2016}]{dauser16}
{Dauser} T.,  {Garc{\'{\i}}a} J.,  {Walton} D.~J.,  {Eikmann} W.,  {Kallman}
  T.,  {McClintock} J.,   {Wilms} J.,  2016, \mn@doi [\aap]
  {10.1051/0004-6361/201628135}, \href
  {http://adsabs.harvard.edu/abs/2016A%26A...590A..76D} {590, A76}

\bibitem[\protect\citeauthoryear{{Di Gesu} et~al.,}{{Di Gesu}
  et~al.}{2013}]{gesu13}
{Di Gesu} L.,  et~al., 2013, \mn@doi [\aap] {10.1051/0004-6361/201321416},
  \href {http://adsabs.harvard.edu/abs/2013A%26A...556A..94D} {556, A94}

\bibitem[\protect\citeauthoryear{{Esin}, {McClintock}  \& {Narayan}}{{Esin}
  et~al.}{1997}]{esin97}
{Esin} A.~A.,  {McClintock} J.~E.,   {Narayan} R.,  1997, \mn@doi [\apj]
  {10.1086/304829}, \href {http://adsabs.harvard.edu/abs/1997ApJ...489..865E}
  {489, 865}

\bibitem[\protect\citeauthoryear{{Fabian} \& {Vaughan}}{{Fabian} \&
  {Vaughan}}{2003}]{reflection3}
{Fabian} A.~C.,  {Vaughan} S.,  2003, \mn@doi [\mnras]
  {10.1046/j.1365-8711.2003.06465.x}, \href
  {http://adsabs.harvard.edu/abs/2003MNRAS.340L..28F} {340, L28}

\bibitem[\protect\citeauthoryear{{Fabian}, {Miniutti}, {Gallo}, {Boller},
  {Tanaka}, {Vaughan}  \& {Ross}}{{Fabian} et~al.}{2004}]{fabian3}
{Fabian} A.~C.,  {Miniutti} G.,  {Gallo} L.,  {Boller} T.,  {Tanaka} Y.,
  {Vaughan} S.,   {Ross} R.~R.,  2004, \mn@doi [\mnras]
  {10.1111/j.1365-2966.2004.08036.x}, \href
  {http://adsabs.harvard.edu/abs/2004MNRAS.353.1071F} {353, 1071}

\bibitem[\protect\citeauthoryear{{Fabian}, {Miniutti}, {Iwasawa}  \&
  {Ross}}{{Fabian} et~al.}{2005}]{fabian05}
{Fabian} A.~C.,  {Miniutti} G.,  {Iwasawa} K.,   {Ross} R.~R.,  2005, \mn@doi
  [\mnras] {10.1111/j.1365-2966.2005.09148.x}, \href
  {http://adsabs.harvard.edu/abs/2005MNRAS.361..795F} {361, 795}

\bibitem[\protect\citeauthoryear{{Fabian} et~al.,}{{Fabian}
  et~al.}{2009}]{1h0707}
{Fabian} A.~C.,  et~al., 2009, \mn@doi [\nat] {10.1038/nature08007}, \href
  {http://adsabs.harvard.edu/abs/2009Natur.459..540F} {459, 540}

\bibitem[\protect\citeauthoryear{{Fabian} et~al.,}{{Fabian}
  et~al.}{2013}]{fabian13}
{Fabian} A.~C.,  et~al., 2013, \mn@doi [\mnras] {10.1093/mnras/sts504}, \href
  {http://adsabs.harvard.edu/abs/2013MNRAS.429.2917F} {429, 2917}

\bibitem[\protect\citeauthoryear{{Fabian}, {Lohfink}, {Belmont}, {Malzac}  \&
  {Coppi}}{{Fabian} et~al.}{2017}]{fabian17}
{Fabian} A.~C.,  {Lohfink} A.,  {Belmont} R.,  {Malzac} J.,   {Coppi} P.,
  2017, \mn@doi [\mnras] {10.1093/mnras/stx221}, \href
  {http://adsabs.harvard.edu/abs/2017MNRAS.467.2566F} {467, 2566}

\bibitem[\protect\citeauthoryear{{Foreman-Mackey}, {Hogg}, {Lang}  \&
  {Goodman}}{{Foreman-Mackey} et~al.}{2013}]{foreman12}
{Foreman-Mackey} D.,  {Hogg} D.~W.,  {Lang} D.,   {Goodman} J.,  2013, \mn@doi
  [\pasp] {10.1086/670067}, \href
  {http://adsabs.harvard.edu/abs/2013PASP..125..306F} {125, 306}

\bibitem[\protect\citeauthoryear{{Gallo} et~al.,}{{Gallo}
  et~al.}{2013}]{gallo13}
{Gallo} L.~C.,  et~al., 2013, \mn@doi [\mnras] {10.1093/mnras/sts102}, \href
  {http://adsabs.harvard.edu/abs/2013MNRAS.428.1191G} {428, 1191}

\bibitem[\protect\citeauthoryear{{Garc{\'{\i}}a}, {Dauser}, {Reynolds},
  {Kallman}, {McClintock}, {Wilms}  \& {Eikmann}}{{Garc{\'{\i}}a}
  et~al.}{2013}]{garcia13}
{Garc{\'{\i}}a} J.,  {Dauser} T.,  {Reynolds} C.~S.,  {Kallman} T.~R.,
  {McClintock} J.~E.,  {Wilms} J.,   {Eikmann} W.,  2013, \mn@doi [\apj]
  {10.1088/0004-637X/768/2/146}, \href
  {http://adsabs.harvard.edu/abs/2013ApJ...768..146G} {768, 146}

\bibitem[\protect\citeauthoryear{{Garc{\'{\i}}a} et~al.,}{{Garc{\'{\i}}a}
  et~al.}{2014}]{garcia14}
{Garc{\'{\i}}a} J.,  et~al., 2014, \mn@doi [\apj] {10.1088/0004-637X/782/2/76},
  \href {http://adsabs.harvard.edu/abs/2014ApJ...782...76G} {782, 76}

\bibitem[\protect\citeauthoryear{{Garc{\'{\i}}a}, {Fabian}, {Kallman},
  {Dauser}, {Parker}, {McClintock}, {Steiner}  \& {Wilms}}{{Garc{\'{\i}}a}
  et~al.}{2016}]{garcia1}
{Garc{\'{\i}}a} J.~A.,  {Fabian} A.~C.,  {Kallman} T.~R.,  {Dauser} T.,
  {Parker} M.~L.,  {McClintock} J.~E.,  {Steiner} J.~F.,   {Wilms} J.,  2016,
  \mn@doi [\mnras] {10.1093/mnras/stw1696}, \href
  {http://adsabs.harvard.edu/abs/2016MNRAS.462..751G} {462, 751}

\bibitem[\protect\citeauthoryear{{Goodman} \& {Weare}}{{Goodman} \&
  {Weare}}{2010}]{goodman10}
{Goodman} J.,  {Weare} J.,  2010, \mn@doi [Communications in Applied
  Mathematics and Computational Science, Vol.~5, No.~1, p.~65-80, 2010]
  {10.2140/camcos.2010.5.65}, \href
  {http://adsabs.harvard.edu/abs/2010CAMCS...5...65G} {5, 65}

\bibitem[\protect\citeauthoryear{{Grupe}, {Komossa}, {Leighly}  \&
  {Page}}{{Grupe} et~al.}{2010}]{grupe10}
{Grupe} D.,  {Komossa} S.,  {Leighly} K.~M.,   {Page} K.~L.,  2010, \mn@doi
  [\apjs] {10.1088/0067-0049/187/1/64}, \href
  {http://adsabs.harvard.edu/abs/2010ApJS..187...64G} {187, 64}

\bibitem[\protect\citeauthoryear{{Guainazzi} et~al.,}{{Guainazzi}
  et~al.}{1998}]{guainazzi98}
{Guainazzi} M.,  et~al., 1998, \aap, \href
  {http://adsabs.harvard.edu/abs/1998A%26A...339..327G} {339, 327}

\bibitem[\protect\citeauthoryear{{Jiang}, {Bambi}  \& {Steiner}}{{Jiang}
  et~al.}{2015}]{jiang15}
{Jiang} J.,  {Bambi} C.,   {Steiner} J.~F.,  2015, \mn@doi [\jcap]
  {10.1088/1475-7516/2015/05/025}, \href
  {http://adsabs.harvard.edu/abs/2015JCAP...05..025J} {5, 025}

\bibitem[\protect\citeauthoryear{{Jiang} et~al.,}{{Jiang}
  et~al.}{2018}]{jiang18}
{Jiang} J.,  et~al., 2018, \mn@doi [\mnras] {10.1093/mnras/sty836}, \href
  {http://adsabs.harvard.edu/abs/2018MNRAS.tmp..804J} {}

\bibitem[\protect\citeauthoryear{{Johannsen} \& {Psaltis}}{{Johannsen} \&
  {Psaltis}}{2010}]{johannsen10}
{Johannsen} T.,  {Psaltis} D.,  2010, \mn@doi [\apj]
  {10.1088/0004-637X/716/1/187}, \href
  {http://adsabs.harvard.edu/abs/2010ApJ...716..187J} {716, 187}

\bibitem[\protect\citeauthoryear{{Kallman} \& {Bautista}}{{Kallman} \&
  {Bautista}}{2001}]{xstar}
{Kallman} T.,  {Bautista} M.,  2001, \mn@doi [\apjs] {10.1086/319184}, \href
  {http://adsabs.harvard.edu/abs/2001ApJS..133..221K} {133, 221}

\bibitem[\protect\citeauthoryear{{Kara}, {Fabian}, {Cackett}, {Steiner},
  {Uttley}, {Wilkins}  \& {Zoghbi}}{{Kara} et~al.}{2013}]{kara13}
{Kara} E.,  {Fabian} A.~C.,  {Cackett} E.~M.,  {Steiner} J.~F.,  {Uttley} P.,
  {Wilkins} D.~R.,   {Zoghbi} A.,  2013, \mn@doi [\mnras]
  {10.1093/mnras/sts155}, \href
  {http://adsabs.harvard.edu/abs/2013MNRAS.428.2795K} {428, 2795}

\bibitem[\protect\citeauthoryear{{Kara}, {Garc{\'{\i}}a}, {Lohfink}, {Fabian},
  {Reynolds}, {Tombesi}  \& {Wilkins}}{{Kara} et~al.}{2017}]{kara17}
{Kara} E.,  {Garc{\'{\i}}a} J.~A.,  {Lohfink} A.,  {Fabian} A.~C.,  {Reynolds}
  C.~S.,  {Tombesi} F.,   {Wilkins} D.~R.,  2017, \mn@doi [\mnras]
  {10.1093/mnras/stx792}, \href
  {http://adsabs.harvard.edu/abs/2017MNRAS.468.3489K} {468, 3489}

\bibitem[\protect\citeauthoryear{Kerr}{Kerr}{1963}]{kerr96}
Kerr R.~P.,  1963, \mn@doi [Phys. Rev. Lett.] {10.1103/PhysRevLett.11.237}, 11,
  237

\bibitem[\protect\citeauthoryear{{Krawczynski} et~al.,}{{Krawczynski}
  et~al.}{2004}]{krawczynski04}
{Krawczynski} H.,  et~al., 2004, \mn@doi [\apj] {10.1086/380393}, \href
  {http://adsabs.harvard.edu/abs/2004ApJ...601..151K} {601, 151}

\bibitem[\protect\citeauthoryear{{Larsson}, {Miniutti}, {Fabian}, {Miller},
  {Reynolds}  \& {Ponti}}{{Larsson} et~al.}{2008}]{larsson08}
{Larsson} J.,  {Miniutti} G.,  {Fabian} A.~C.,  {Miller} J.~M.,  {Reynolds}
  C.~S.,   {Ponti} G.,  2008, \mn@doi [\mnras]
  {10.1111/j.1365-2966.2008.12844.x}, \href
  {http://adsabs.harvard.edu/abs/2008MNRAS.384.1316L} {384, 1316}

\bibitem[\protect\citeauthoryear{{Marinucci} et~al.,}{{Marinucci}
  et~al.}{2014a}]{marinucci14}
{Marinucci} A.,  et~al., 2014a, \mn@doi [\mnras] {10.1093/mnras/stu404}, \href
  {http://adsabs.harvard.edu/abs/2014MNRAS.440.2347M} {440, 2347}

\bibitem[\protect\citeauthoryear{{Marinucci} et~al.,}{{Marinucci}
  et~al.}{2014b}]{marinucci}
{Marinucci} A.,  et~al., 2014b, \mn@doi [\apj] {10.1088/0004-637X/787/1/83},
  \href {http://adsabs.harvard.edu/abs/2014ApJ...787...83M} {787, 83}

\bibitem[\protect\citeauthoryear{{McHardy}, {Koerding}, {Knigge}, {Uttley}  \&
  {Fender}}{{McHardy} et~al.}{2006}]{mchardy06}
{McHardy} I.~M.,  {Koerding} E.,  {Knigge} C.,  {Uttley} P.,   {Fender} R.~P.,
  2006, \mn@doi [\nat] {10.1038/nature05389}, \href
  {http://adsabs.harvard.edu/abs/2006Natur.444..730M} {444, 730}

\bibitem[\protect\citeauthoryear{{Miniutti} \& {Fabian}}{{Miniutti} \&
  {Fabian}}{2004}]{miniutti04}
{Miniutti} G.,  {Fabian} A.~C.,  2004, \mn@doi [\mnras]
  {10.1111/j.1365-2966.2004.07611.x}, \href
  {http://adsabs.harvard.edu/abs/2004MNRAS.349.1435M} {349, 1435}

\bibitem[\protect\citeauthoryear{{Miniutti}, {Panessa}, {de Rosa}, {Fabian},
  {Malizia}, {Molina}, {Miller}  \& {Vaughan}}{{Miniutti}
  et~al.}{2009}]{miniutti09}
{Miniutti} G.,  {Panessa} F.,  {de Rosa} A.,  {Fabian} A.~C.,  {Malizia} A.,
  {Molina} M.,  {Miller} J.~M.,   {Vaughan} S.,  2009, \mn@doi [\mnras]
  {10.1111/j.1365-2966.2009.15092.x}, \href
  {http://adsabs.harvard.edu/abs/2009MNRAS.398..255M} {398, 255}

\bibitem[\protect\citeauthoryear{{Narayan}}{{Narayan}}{2005}]{narayan05}
{Narayan} R.,  2005, \mn@doi [\apss] {10.1007/s10509-005-1178-7}, \href
  {http://adsabs.harvard.edu/abs/2005Ap%26SS.300..177N} {300, 177}

\bibitem[\protect\citeauthoryear{{Oh} et~al.,}{{Oh} et~al.}{2018}]{oh18}
{Oh} K.,  et~al., 2018, \mn@doi [\apjs] {10.3847/1538-4365/aaa7fd}, \href
  {http://adsabs.harvard.edu/abs/2018ApJS..235....4O} {235, 4}

\bibitem[\protect\citeauthoryear{{Parker} et~al.,}{{Parker}
  et~al.}{2014}]{parker14}
{Parker} M.~L.,  et~al., 2014, \mn@doi [\mnras] {10.1093/mnras/stu1246}, \href
  {http://adsabs.harvard.edu/abs/2014MNRAS.443.1723P} {443, 1723}

\bibitem[\protect\citeauthoryear{{Parker} et~al.,}{{Parker}
  et~al.}{2015}]{parker15}
{Parker} M.~L.,  et~al., 2015, \mn@doi [\mnras] {10.1093/mnras/stu2424}, \href
  {http://adsabs.harvard.edu/abs/2015MNRAS.447...72P} {447, 72}

\bibitem[\protect\citeauthoryear{{Pounds}}{{Pounds}}{2005}]{pounds05}
{Pounds} K.,  2005, in {Briel} U.~G.,  {Sembay} S.,   {Read} A.,  eds, 5 years
  of Science with XMM-Newton. pp 65--71 (\mn@eprint {} {astro-ph/0505447})

\bibitem[\protect\citeauthoryear{{Pounds}, {Reeves}, {Page}  \&
  {O'Brien}}{{Pounds} et~al.}{2004a}]{pounds04a}
{Pounds} K.~A.,  {Reeves} J.~N.,  {Page} K.~L.,   {O'Brien} P.~T.,  2004a,
  \mn@doi [\apj] {10.1086/382678}, \href
  {http://adsabs.harvard.edu/abs/2004ApJ...605..670P} {605, 670}

\bibitem[\protect\citeauthoryear{{Pounds}, {Reeves}, {Page}  \&
  {O'Brien}}{{Pounds} et~al.}{2004b}]{pounds04b}
{Pounds} K.~A.,  {Reeves} J.~N.,  {Page} K.~L.,   {O'Brien} P.~T.,  2004b,
  \mn@doi [\apj] {10.1086/424992}, \href
  {http://adsabs.harvard.edu/abs/2004ApJ...616..696P} {616, 696}

\bibitem[\protect\citeauthoryear{{Reis} et~al.,}{{Reis} et~al.}{2012}]{reis12}
{Reis} R.~C.,  et~al., 2012, \mn@doi [\apj] {10.1088/0004-637X/745/1/93}, \href
  {http://adsabs.harvard.edu/abs/2012ApJ...745...93R} {745, 93}

\bibitem[\protect\citeauthoryear{{Reynolds}}{{Reynolds}}{2014}]{reynolds14}
{Reynolds} C.~S.,  2014, \mn@doi [\ssr] {10.1007/s11214-013-0006-6}, \href
  {http://adsabs.harvard.edu/abs/2014SSRv..183..277R} {183, 277}

\bibitem[\protect\citeauthoryear{{Ricci}, {Tazaki}, {Ueda}, {Paltani},
  {Boissay}  \& {Terashima}}{{Ricci} et~al.}{2014}]{ricci14}
{Ricci} C.,  {Tazaki} F.,  {Ueda} Y.,  {Paltani} S.,  {Boissay} R.,
  {Terashima} Y.,  2014, \mn@doi [\apj] {10.1088/0004-637X/795/2/147}, \href
  {http://adsabs.harvard.edu/abs/2014ApJ...795..147R} {795, 147}

\bibitem[\protect\citeauthoryear{{Risaliti} et~al.,}{{Risaliti}
  et~al.}{2013}]{risaliti13}
{Risaliti} G.,  et~al., 2013, \mn@doi [\nat] {10.1038/nature11938}, \href
  {http://adsabs.harvard.edu/abs/2013Natur.494..449R} {494, 449}

\bibitem[\protect\citeauthoryear{{Sesana}, {Barausse}, {Dotti}  \&
  {Rossi}}{{Sesana} et~al.}{2014}]{senana14}
{Sesana} A.,  {Barausse} E.,  {Dotti} M.,   {Rossi} E.~M.,  2014, \mn@doi
  [\apj] {10.1088/0004-637X/794/2/104}, \href
  {http://adsabs.harvard.edu/abs/2014ApJ...794..104S} {794, 104}

\bibitem[\protect\citeauthoryear{{Shemmer}, {Brandt}, {Netzer}, {Maiolino}  \&
  {Kaspi}}{{Shemmer} et~al.}{2006}]{shemmer06}
{Shemmer} O.,  {Brandt} W.~N.,  {Netzer} H.,  {Maiolino} R.,   {Kaspi} S.,
  2006, \mn@doi [\apjl] {10.1086/506911}, \href
  {http://adsabs.harvard.edu/abs/2006ApJ...646L..29S} {646, L29}

\bibitem[\protect\citeauthoryear{{Svensson} \& {Zdziarski}}{{Svensson} \&
  {Zdziarski}}{1994}]{svensson94}
{Svensson} R.,  {Zdziarski} A.~A.,  1994, \mn@doi [\apj] {10.1086/174934},
  \href {http://adsabs.harvard.edu/abs/1994ApJ...436..599S} {436, 599}

\bibitem[\protect\citeauthoryear{{Tan}, {Wang}, {Shu}  \& {Zhou}}{{Tan}
  et~al.}{2012}]{tan12}
{Tan} Y.,  {Wang} J.~X.,  {Shu} X.~W.,   {Zhou} Y.,  2012, \mn@doi [\apjl]
  {10.1088/2041-8205/747/1/L11}, \href
  {http://adsabs.harvard.edu/abs/2012ApJ...747L..11T} {747, L11}

\bibitem[\protect\citeauthoryear{{Tanaka} et~al.,}{{Tanaka}
  et~al.}{1995}]{tanaka95}
{Tanaka} Y.,  et~al., 1995, \mn@doi [\nat] {10.1038/375659a0}, \href
  {http://adsabs.harvard.edu/abs/1995Natur.375..659T} {375, 659}

\bibitem[\protect\citeauthoryear{{Tazaki}, {Ueda}, {Ishino}, {Eguchi}, {Isobe},
  {Terashima}  \& {Mushotzky}}{{Tazaki} et~al.}{2010}]{tazaki10}
{Tazaki} F.,  {Ueda} Y.,  {Ishino} Y.,  {Eguchi} S.,  {Isobe} N.,  {Terashima}
  Y.,   {Mushotzky} R.~F.,  2010, \mn@doi [\apj]
  {10.1088/0004-637X/721/2/1340}, \href
  {http://adsabs.harvard.edu/abs/2010ApJ...721.1340T} {721, 1340}

\bibitem[\protect\citeauthoryear{{Thomas}, {Beuermann}, {Reinsch}, {Schwope},
  {Truemper}  \& {Voges}}{{Thomas} et~al.}{1998}]{thomas98}
{Thomas} H.-C.,  {Beuermann} K.,  {Reinsch} K.,  {Schwope} A.~D.,  {Truemper}
  J.,   {Voges} W.,  1998, \aap, \href
  {http://adsabs.harvard.edu/abs/1998A%26A...335..467T} {335, 467}

\bibitem[\protect\citeauthoryear{{Tombesi}, {Cappi}, {Reeves}, {Palumbo},
  {Yaqoob}, {Braito}  \& {Dadina}}{{Tombesi} et~al.}{2010}]{tombesi1}
{Tombesi} F.,  {Cappi} M.,  {Reeves} J.~N.,  {Palumbo} G.~G.~C.,  {Yaqoob} T.,
  {Braito} V.,   {Dadina} M.,  2010, \mn@doi [\aap]
  {10.1051/0004-6361/200913440}, \href
  {http://adsabs.harvard.edu/abs/2010A%26A...521A..57T} {521, A57}

\bibitem[\protect\citeauthoryear{{Tomsick} et~al.,}{{Tomsick}
  et~al.}{2018}]{tomsick18}
{Tomsick} J.~A.,  et~al., 2018, \mn@doi [\apj] {10.3847/1538-4357/aaaab1},
  \href {http://adsabs.harvard.edu/abs/2018ApJ...855....3T} {855, 3}

\bibitem[\protect\citeauthoryear{{Tortosa} et~al.,}{{Tortosa}
  et~al.}{2017}]{tortosa17}
{Tortosa} A.,  et~al., 2017, \mn@doi [\mnras] {10.1093/mnras/stw3301}, \href
  {http://adsabs.harvard.edu/abs/2017MNRAS.466.4193T} {466, 4193}

\bibitem[\protect\citeauthoryear{{Turner}, {Reeves}, {Braito}  \&
  {Costa}}{{Turner} et~al.}{2018}]{turner18}
{Turner} T.~J.,  {Reeves} J.~N.,  {Braito} V.,   {Costa} M.,  2018, \mn@doi
  [\mnras] {10.1093/mnras/sty318}, \href
  {http://adsabs.harvard.edu/abs/2018MNRAS.476.1258T} {476, 1258}

\bibitem[\protect\citeauthoryear{{Vasudevan} \& {Fabian}}{{Vasudevan} \&
  {Fabian}}{2007}]{bolometric}
{Vasudevan} R.~V.,  {Fabian} A.~C.,  2007, \mn@doi [\mnras]
  {10.1111/j.1365-2966.2007.12328.x}, \href
  {http://adsabs.harvard.edu/abs/2007MNRAS.381.1235V} {381, 1235}

\bibitem[\protect\citeauthoryear{{Walton}, {Reis}, {Cackett}, {Fabian}  \&
  {Miller}}{{Walton} et~al.}{2012}]{walton12b}
{Walton} D.~J.,  {Reis} R.~C.,  {Cackett} E.~M.,  {Fabian} A.~C.,   {Miller}
  J.~M.,  2012, \mn@doi [\mnras] {10.1111/j.1365-2966.2012.20809.x}, \href
  {http://adsabs.harvard.edu/abs/2012MNRAS.422.2510W} {422, 2510}

\bibitem[\protect\citeauthoryear{{Walton}, {Nardini}, {Fabian}, {Gallo}  \&
  {Reis}}{{Walton} et~al.}{2013}]{walton13}
{Walton} D.~J.,  {Nardini} E.,  {Fabian} A.~C.,  {Gallo} L.~C.,   {Reis} R.~C.,
   2013, \mn@doi [\mnras] {10.1093/mnras/sts227}, \href
  {http://adsabs.harvard.edu/abs/2013MNRAS.428.2901W} {428, 2901}

\bibitem[\protect\citeauthoryear{{Walton} et~al.,}{{Walton}
  et~al.}{2014}]{walton14}
{Walton} D.~J.,  et~al., 2014, \mn@doi [\apj] {10.1088/0004-637X/788/1/76},
  \href {http://adsabs.harvard.edu/abs/2014ApJ...788...76W} {788, 76}

\bibitem[\protect\citeauthoryear{{Walton} et~al.,}{{Walton}
  et~al.}{2018}]{walton18b}
{Walton} D.~J.,  et~al., 2018, \mn@doi [\mnras] {10.1093/mnras/stx2659}, \href
  {http://adsabs.harvard.edu/abs/2018MNRAS.473.4377W} {473, 4377}

\bibitem[\protect\citeauthoryear{{Willingale}, {Starling}, {Beardmore},
  {Tanvir}  \& {O'Brien}}{{Willingale} et~al.}{2013}]{willingale13}
{Willingale} R.,  {Starling} R.~L.~C.,  {Beardmore} A.~P.,  {Tanvir} N.~R.,
  {O'Brien} P.~T.,  2013, \mn@doi [\mnras] {10.1093/mnras/stt175}, \href
  {http://adsabs.harvard.edu/abs/2013MNRAS.431..394W} {431, 394}

\bibitem[\protect\citeauthoryear{{Wilms}, {Allen}  \& {McCray}}{{Wilms}
  et~al.}{2000}]{wilms}
{Wilms} J.,  {Allen} A.,   {McCray} R.,  2000, \mn@doi [\apj] {10.1086/317016},
  \href {http://adsabs.harvard.edu/abs/2000ApJ...542..914W} {542, 914}

\bibitem[\protect\citeauthoryear{{Wilms}, {Reynolds}, {Begelman}, {Reeves},
  {Molendi}, {Staubert}  \& {Kendziorra}}{{Wilms} et~al.}{2001}]{wilms01}
{Wilms} J.,  {Reynolds} C.~S.,  {Begelman} M.~C.,  {Reeves} J.,  {Molendi} S.,
  {Staubert} R.,   {Kendziorra} E.,  2001, \mn@doi [\mnras]
  {10.1046/j.1365-8711.2001.05066.x}, \href
  {http://adsabs.harvard.edu/abs/2001MNRAS.328L..27W} {328, L27}

\bibitem[\protect\citeauthoryear{Yan, Sadeghpour  \& Dalgarno}{Yan
  et~al.}{2001}]{crosssec2}
Yan M.,  Sadeghpour H.~R.,   Dalgarno A.,  2001, The Astrophysical Journal,
  559, 1194

\bibitem[\protect\citeauthoryear{{Zdziarski}, {Johnson}  \&
  {Magdziarz}}{{Zdziarski} et~al.}{1996}]{zdziarski99}
{Zdziarski} A.~A.,  {Johnson} W.~N.,   {Magdziarz} P.,  1996, \mn@doi [\mnras]
  {10.1093/mnras/283.1.193}, \href
  {http://adsabs.harvard.edu/abs/1996MNRAS.283..193Z} {283, 193}

\bibitem[\protect\citeauthoryear{{Zhang}, {Treves}, {Maraschi}, {Bai}  \&
  {Liu}}{{Zhang} et~al.}{2006}]{zhang06}
{Zhang} Y.~H.,  {Treves} A.,  {Maraschi} L.,  {Bai} J.~M.,   {Liu} F.~K.,
  2006, \mn@doi [\apj] {10.1086/498498}, \href
  {http://adsabs.harvard.edu/abs/2006ApJ...637..699Z} {637, 699}

\bibitem[\protect\citeauthoryear{{{\.Z}ycki}, {Done}  \& {Smith}}{{{\.Z}ycki}
  et~al.}{1999}]{zycki99}
{{\.Z}ycki} P.~T.,  {Done} C.,   {Smith} D.~A.,  1999, \mn@doi [\mnras]
  {10.1046/j.1365-8711.1999.02885.x}, \href
  {http://adsabs.harvard.edu/abs/1999MNRAS.309..561Z} {309, 561}

\makeatother
\end{thebibliography}







\bsp	
\label{lastpage}
\end{document}